\documentclass[11pt, oneside]{article}   	
\usepackage{geometry,amsmath,amssymb,amsthm,amsfonts,mathrsfs,marvosym,graphicx,graphics,keyval,young}
\usepackage{braket,ytableau}  

\usepackage[toc,page]{appendix}
\usepackage{graphicx}				
\usepackage{amssymb}

\newcommand{\mc}{\mathcal}
\newcommand{\mrn}{\mathrm}

\title{\bf{\huge{Restricted Schurs and correlators for $SO(N)$ and $Sp(N)$}}}
\author{\bf{\Large{Garreth Kemp$^{1}$\footnote{garreth.kemp@students.wits.ac.za}}}}
\date{}			
\begin{document}

\begin{titlepage}
\maketitle

\begin{center}
	\emph{$^{1}$National Institute for Theoretical Physics,\\
		School of Physics and Centre for Theoretical Physics,\\
		University of the Witwatersrand, Wits, 2050,\\
		South Africa}
\end{center}

\begin{abstract}

In a recent work, restricted Schur polynomials have been argued to form a complete orthogonal set of gauge invariant operators for the $1/4$-BPS sector of free $\mc{N} = 4$ super Yang-Mills theory with an $SO(N)$ gauge group. In this work, we extend these results to the theory with an $Sp(N)$ gauge group. Using these operators, we develop techniques to compute correlation functions of any multi-trace operators with two scalar fields exactly in the free theory limit for both $SO(N)$ and $Sp(N)$.

\end{abstract}
\end{titlepage}
\tableofcontents


\section{Introduction}

Restricted Schur polynomials have been argued to form a complete orthogonal set of gauge invariant operators for the $1/4$-BPS sector of free $\mc{N} = 4$ super Yang-Mills theory with an $SO(N)$ gauge group \cite{Kemp:2014apa}. For even $N$, \cite{Kemp:2014apa} employed representation theory of the symmetric and hyper octahedral groups to construct a complete set of local operators depending on two scalar fields. Their two-point function was computed exactly and shown to be diagonal in the labels of these operators. More specifically, these operators were the generalisations of the so-called restricted Schur polynomials of the $U(N)$ gauge theory \cite{GGWSA1}, \cite{GGWSA2} and \cite{GGWSA3}.

Schur polynomials $\chi_{R}(Z)$ in the 1/2-BPS sector of the free $U(N)$ gauge theory were first studied in \cite{Jevicki}, in which these operators, labeled by irreducible representations (irreps) $R$ of the symmetric and unitary groups, were shown to be an exactly orthogonal basis. Thus, these gauge invariant operators could be used as a basis to study the large $N$ limit of operators whose dimension scales parametrically with $N$. The trace basis, for example, is no longer orthogonal in this case and computing non-planar corrections in correlation functions is a difficult task. 
When the Young diagram $R$ labelling the Schur polynomial has long columns, or long rows, ($O(N)$ boxes in each row/column) the operator is dual to a system of giant gravitons moving in the $S^{5}$ or AdS$_{5}$ \cite{Invasion}, \cite{Hashimoto:2000zp}, \cite{Meyers}, \cite{Jevicki} and \cite{Emergent}.

By adding a different type of field to $\chi_{R}(Z)$, one can build the restricted Schur polynomial. A restricted Schur may simply be thought of as a linear combination all possible multi-matrix, multi-trace operators, where the sum is over permutations of the symmetric group. We can add different types of complex scalar fields, fermion fields or even gauge fields \cite{XYZmatrices}, \cite{Fermions} and \cite{deMelloKoch:2011vn}. For two scalar fields, the counting of restricted Schurs was shown to match the counting of states of the free theory in \cite{Collins}. Their two-point function was computed exactly in \cite{EMC} and was shown to be diagonal in the operator labels. \cite{Bhattacharyya:2008xy} succeeded in writing any multi-trace operator involving two scalar fields as a linear combination of restricted Schurs. They also derived a product rule for these operators. This allowed the product of any two restricted Schurs to be written in terms of restricted Schurs and restricted Littlewood-Richardson coefficients.

At the level of summing Feynman diagrams, computing correlation functions of multi-trace multi-matrix operators is a difficult task at finite $N$. This is because the non-planar corrections are no longer suppressed and must be taken into account. The results of \cite{EMC} and \cite{Bhattacharyya:2008xy} transformed the problem of computing correlation functions of the multi-matrix, multi-trace operators into the problem of computing correlation functions of restricted Schur polynomials - something we can do exactly for any $N$.

Other orthogonal bases for the 1/4 and 1/8-BPS sectors of the free gauge theory has been proposed. \cite{Kimura:2007wy} constructed operators from $Z$ and $Z^{\dagger}$, while \cite{Brown:2007xh} constructed operators depending on $X,Y$ and $Z$. At weak coupling some progress towards understanding the anomalous dimensions of an orthogonal basis of the free theory, which is manifestly covariant under the global symmetry, was achieved in \cite{Weakcoup}.

Restricted Schur polynomials with $n$ $Z$'s and $m$ $Y$'s are labeled by three Young diagram labels, $R,r,s$, corresponding to irreducible representations of $S_{n+m}$ and $S_{n}\times S_{m}$ respectively. When the number of $Z$'s and $Y$'s is $O(N)$, with $n \gg m$, these operators again have a D-brane interpretation in the string theory. For $R$ having long columns (or rows), each with $O(N)$ boxes, the operator is dual to excited giant gravitons \cite{GGWSA1}. A system of excited giant gravitons can be thought of as a system of giants with strings attached. Amongst the other bases found for the 1/4-BPS sector it has been argued that restricted Schur polynomials is the most natural basis for studying open string dynamics of their dual $D$-brane states \cite{Bhattacharyya:2008xy}. To this end, the one-loop dilation operator has been diagonalised and the spectrum of anomalous dimensions computed in \cite{Koch:2010gp}, \cite{DeComarmond:2010ie}, \cite{Carlson:2011hy}, \cite{GGO} \cite{Spring}, \cite{deMelloKoch:2012ck}. The two-loop case was studied in \cite{twoloop}. Remarkably, for the non-planar limit studied in these works the spectrum was shown to be that of a system of decouple harmonic oscillators. This is evidence of integrability in these non-planar sectors of the gauge theory. 

A similar program has been initiated for the gauge theory with an orthogonal gauge group in \cite{SON1}, and \cite{SON2}. In the 1/2-BPS sector, Schur polynomials were constructed from the basic building block $\ytableausetup{boxsize=0.55em}\ydiagram[]{2,2}$ which diagonalised the two-point function in the free theory limit. Indeed, orthogonality of Schur polynomials built from a single scalar field $Z$ is a gauge group-independent property \cite{Diaz}. The counting of states, as given by the partition function, for $SO(4)$ and $SO(6)$, was shown to match the number of Schur polynomials that could be defined. This basis was related to the trace basis and a product rule was derived for the Schurs. These results were then extended to the theory with symplectic gauge group, $Sp(N)$. $Sp(N)$ is related to $SO(N)$ by exchanging symmetrisations of irreps and replacing $N$ by $-N$ \cite{SON2}, \cite{SpecProbSON}.

This study then progressed to the 1/4-BPS sector of the free $SO(N)$ gauge theory \cite{Kemp:2014apa}. The counting of states, as given by the partition function, and the counting of restricted Schur polynomials was shown to agree by restricting to a particular class of Young diagram labels. $R$ must have an even number of boxes in each row, while $r$ and $s$ are restricted to have an even number of boxes in each column. An explicit construction of these operators was given and their two-point function was evaluated exactly and shown to be diagonal.

Physics in the non-planar limit of $SO(N)$ and $Sp(N)$ gauge theory is different from that of $U(N)$. The $SO(N)$ and $Sp(N)$ gauge theories are matrix models with anti-symmetric and symplectic matrices respectively. When evaluating correlation functions, the leading non-planar correction comes from non-orientable Feynman diagrams with a single cross-cap - an effect not present in the $U(N)$ theory \cite{SON2}, \cite{SpecProbSON}, \cite{Cicuta:1982fu}. Furthermore, $\mc{N} = 4$ super Yang-Mills theory with an $SO(N)$ or $Sp(N)$ gauge group is dual to type IIB string theory with AdS$_{5}\times \mc{R}P^{5}$ geometry. The Schur polynomials of \cite{SON1}, \cite{SON2} and the restricted Schur polynomials of \cite{Kemp:2014apa} may prove useful as a basis of operators to study non-planar limits of these gauge and dual string theories.

In this paper, we extend our results found in \cite{Kemp:2014apa} for $SO(N)$ to $Sp(N)$. First, we express the free 1/4-BPS $Sp(N)$ partition function in terms the Littlewood-Richardson coefficients which count the number of $Sp(N)$ restricted Schurs. We then give a gauge invariant construction of these operators and evaluate their two-point function. As expected, the results are identical to those for $SO(N)$ except the Young diagrams are transposed and $N$ is replaced by $-N$.  We then relate the trace basis to the restricted Schur basis for both gauge groups. Lastly, we derive a product rule for our operators.


\section{Recap of $SO(N)$}

With a more convenient normalisation, the restricted Schurs defined in \cite{Kemp:2014apa} were

\begin{equation}
\label{eq:2}
	O^{SO(N)}_{R(r,s)\alpha}(Z,Y) = \frac{1}{\sqrt{n!m!q!2^{2q}}}\sum\limits_{\sigma \in S_{2q}}\chi^{SO(N)}_{R(r,s)\alpha}(\sigma) C^{4\nu}_{I} \sigma^{I}_{J}(Z^{\otimes 2n}\otimes Y^{\otimes 2m})^{J}
\end{equation}
\\
where we defined the $SO(N)$ restricted character to be

\begin{eqnarray}
\label{eq:9}
	\chi^{SO(N)}_{R(r,s)\alpha}(\sigma)  = \mrn{Tr}\big( \mc{O}^{SO(N)}_{R(r,s)\alpha}\Gamma_{R}(\sigma) \big), \hspace{20pt} \mrn{and} \hspace{20pt} \mc{O}^{SO(N)}_{R(r,s)\alpha} = \ket{[S]}\!\!\bra{[A]_{r},[A]_{s},\beta}P_{R\rightarrow (r,s)\beta\alpha}
\end{eqnarray}
\\
In (\ref{eq:9}), we have explicitly indicated which irreps subduce the two $[A]$'s. Irrep $r$ of $S_{2n}$ subduces irrep $[A]_{r}$ of $S_{n}[S_{2}]$, and irrep $s$ of $S_{2m}$ subduces irrep $[A]_{s}$ of $S_{m}[S_{2}]$. $\beta$ still labels the particular copy of $(r,s)$ subduced from irrep $R$ of $S_{2q}$. Recall that the state $\ket{[A]_{r},[A]_{s},\beta}$ spans the 1-dimensional carrier space of the $S_{n}[S_{2}]\times S_{m}[S_{2}]$ irrep $([A]_{r},[A]_{s})\beta$, and is calculated as the eigenvector of $P_{R\rightarrow ([A]_{r},[A]_{s})\beta}$ with eigenvalue 1. The intertwiner $P_{R\rightarrow(r,s)\alpha\beta}$ maps this state from one copy of $(r,s)$ to another

\begin{equation}
\label{eq:10}
	P_{R\rightarrow(r,s)\alpha\beta} \ket{[A]_{r},[A]_{s},\beta} =  \ket{[A]_{r},[A]_{s},\alpha}
\end{equation}
\\
Thus, we can write

\begin{equation}
\label{eq:11a}
	\mc{O}^{SO(N)}_{R(r,s)\alpha} = \ket{[S]}\!\!\bra{[A]_{r},[A]_{s},\alpha}.
\end{equation}
\\
For the normalisation in (\ref{eq:2}), the two-point function is

\begin{equation}
\label{eq:27}
	\langle O^{SO(N)}_{R(r,s)\alpha}(Z,Y)\overline{O}^{SO(N)}_{T(t,u)\beta}(Z,Y) \rangle = \delta_{RT}\delta_{rt}\delta_{su}\delta_{\alpha\beta}\frac{(2q)!}{d_{R}}\prod_{i\;\in\;\mrn{odd\;columns}}c_{i}
\end{equation}
\\
The tensor 

\begin{equation}
\label{eq:Contractor}
	C^{4\nu}_{I} = \delta_{k_{1}k_{2}} \cdots \delta_{k_{2q-1}k_{2q}}(\sigma_{4\nu})^{K}_{I} = \delta_{K}(\sigma_{4\nu})^{K}_{I}
\end{equation}
\\
contracts the free indices in such a way as to produce a gauge invariant operator. In \cite{Kemp:2014apa} we chose $\sigma_{4\nu} = (1,2,3,4)\cdots (2q-3,2q-2,2q-1,2q)$. This contracted indices $1\&4, 2\&3 \cdots 2q-3\&2q$ and $2q-2\&2q-1$. The permutation $\sigma_{4\nu} = (1,2)(3,4)\cdots (2q-1,2q)$, or simply the identity permutation, gives the same gauge invariant operator. Thus, an equivalent definition for $O_{R(r,s)\alpha}(Z,Y)$ is\footnote{The $\ket{[S]}$ in this operator is symmetric in boxes $1\&2$, $3\&4\cdots$, $2q-1\&2q$.}

\begin{equation}
\label{eq:Newdefintermsofdelta}
	O^{SO(N)}_{R(r,s)\alpha}(Z,Y) = \frac{1}{\sqrt{n!m!q!2^{2q}}}\sum\limits_{\sigma \in S_{2q}} \chi^{SO(N)}_{R(r,s)\alpha}(\sigma)\,\delta_{I}\sigma^{I}_{J}(Z^{\otimes 2n}\otimes Y^{\otimes 2m})^{J}
\end{equation}
 \\ 
For the permutations $\sigma \in S_{2q}$, $\delta_{I}\sigma^{I}_{J}(Z^{\otimes 2n}\otimes Y^{\otimes 2m})^{J}$ gives all the possible multi-trace operators involving the two scalar fields $Z$ and $Y$. For example, consider $q=4$ with $n=m=2$. There are only 4 multi-trace operators we can define. Here they are for 4 examples of $\sigma$

\begin{eqnarray}
\label{eq:example1}
	\sigma = (2,4,6,8,3) \hspace{10pt} &\mrn{gives}& \delta_{I}\sigma^{I}_{J}(Z^{\otimes 4}\otimes Y^{\otimes 4})^{J} = \mrn{Tr}(Z^{2}Y^{2})\\
\label{eq:example2}
	\sigma = (2,5)(3,4,5) \hspace{10pt} &\mrn{gives}& \delta_{I}\sigma^{I}_{J}(Z^{\otimes 4}\otimes Y^{\otimes 4})^{J} = \mrn{Tr}(ZY)^{2}\\
	\label{eq:example3}
	\sigma = (1,3,2)(5,8,7) \hspace{10pt} &\mrn{gives}& \delta_{I}\sigma^{I}_{J}(Z^{\otimes 4}\otimes Y^{\otimes 4})^{J} = \mrn{Tr}(Z^{2})\mrn{Tr}(Y^{2})\\
	\label{eq:example4}
	\sigma = (1,5,2)(3,4,8,6,7) \hspace{10pt} &\mrn{gives}& \delta_{I}\sigma^{I}_{J}(Z^{\otimes 4}\otimes Y^{\otimes 4})^{J} = \mrn{Tr}(ZY\!ZY)
\end{eqnarray}


\section{Counting for $Sp(N)$}

The partition function for the $1/4$-BPS sector of free $Sp(N)$ gauge theory is \cite{Dolan2}, \cite{Aharony:2003sx}

\begin{equation}
\label{eq:2.1}
	G_{Sp(N)}(t_{1},t_{2}) = \int_{O\in Sp(N)} [dO]e^{\sum^{\infty}_{m=1}\left(\frac{t^{m}_{1}+t^{m}_{2}}{m}\right)\chi_{\mrn{adj}}(O^{m})}
\end{equation}
\\
where $\chi_{\mrn{adj}}(O)$ is the character of $O$ in the adjoint representation of $Sp(N)$, and $[dO]$ is the $Sp(N)$-invariant measure. We take $N=2n$. In terms of the eigenvalues of $O$, the adjoint character  and the integration measure are \cite{Dolan1}

\begin{eqnarray}
\label{eq:2.2}
	\chi^{\mrn{adj}}_{Sp(N)}(\mrn{x}) = \sum\limits_{1\leqslant i < j \leqslant n} (x_{i}x_{j}+x^{-1}_{i}x_{j}+x_{i}x^{-1}_{j}+x^{-1}_{i}x^{-1}_{j}) + \sum\limits^{n}_{i=1}(x^{2}_{i}+x^{-2}_{i}) + n
\end{eqnarray}
\\
and

\begin{eqnarray}
\label{eq:2.3}
	 \hspace{20pt} \int_{O\in Sp(N)}[dO] f(\mrn{x}) &=& \frac{(-1)^{n}}{2^{n}n!}\int_{T_{n}} \prod\limits^{n}_{j=1}\frac{dx_{j}}{2\pi x_{j}} \prod\limits^{n}_{j=1}(x_{j}-x^{-1}_{j})^{2}\Delta(\mrn{x}+\mrn{x}^{-1})^{2}f(\mrn{x})
\end{eqnarray}
\\
where $T_{n} = S^{1}\times S^{1} \cdots \times S^{1}$ and $f(\mrn{x}), \mrn{x} = (x_{1}, x_{2}\cdots x_{n})$, is any symmetric function. Using (\ref{eq:2.2}) the exponential in (\ref{eq:2.1}), after some algebra, becomes

\begin{eqnarray}
\label{2.4}
	e^{\sum^{\infty}_{m=1}\left(\frac{t^{m}_{1}+t^{m}_{2}}{m}\right)\chi_{\mrn{adj}}(O^{m})} \!\!\!\! &=& \!\!\!\!e^{\sum^{\infty}_{m=1}\left( \frac{t^{m}_{1}+t^{m}_{2}}{m} \right)\sum_{1\leqslant i < j \leqslant n} (x^{m}_{i}x^{m}_{j}+x^{-m}_{i}x^{m}_{j}+x_{i}x^{-m}_{j}+x^{-m}_{i}x^{-m}_{j}) + \sum^{n}_{i=1}(x^{2m}_{i}+x^{-2m}_{i}) + n} \nonumber \\
	&=& \!\!\!\!\prod\limits^{2}_{k=1}\prod\limits_{1\leqslant i< j\leqslant n}\frac{1}{(1-t_{k}x_{i}x_{j})(1-t_{k}x^{-1}_{i}x_{j})(1-t_{k}x_{i}x^{-1}_{j})(1-t_{k}x^{-1}_{i}x^{-1}_{j})}\nonumber\\
\label{eq:2.5}
	&& \hspace{20pt} \times \; \frac{1}{(1-t_{k})^{n}}\prod\limits_{1\leqslant i \leqslant n}\frac{1}{(1-t_{k}x^{2}_{i})(1-t_{k}x^{-2}_{i})}
\end{eqnarray}
\\
Changing variables

\begin{eqnarray}
\label{eq:2.6}
	y_{i} &=& x_{\lceil i/2 \rceil} \hspace{20pt} \mrn{for\;odd\;}i\\
\label{2.7}
	y_{2i} &=& x^{-1}_{i} \hspace{20pt} \mrn{for\;even\;}i
\end{eqnarray}
\\
the exponential becomes

\begin{eqnarray}
\label{eq:2.8}
	e^{\sum^{\infty}_{m=1}\left(\frac{t^{m}_{1}+t^{m}_{2}}{m}\right)\chi_{\mrn{adj}}(O^{m})} &=& \prod\limits^{2}_{k=1}\prod\limits_{1\leqslant i \leqslant j \leqslant N} \frac{1}{1-t_{k}y_{i}y_{j}}\\
\label{eq:2.9}
	&=&  \prod\limits^{2}_{k=1}\left( \sum\limits_{\lambda}s_{2\lambda}(\sqrt{t_{k}}y_{1},\sqrt{t_{k}}y_{2}\cdots \sqrt{t_{k}}y_{N}) \right)\\
\label{eq:2.10}
	&=& \sum\limits_{\mu}\sum\limits_{\lambda} (t_{1})^{|2\lambda|/2}(t_{2})^{|2\mu|/2}s_{2\lambda}(y_{1},\cdots y_{N})s_{2\mu}(y_{1},\cdots, y_{N})
\end{eqnarray}
\\
The expansion in terms of Schur functions we used in (\ref{eq:2.9}) is an identity found in \cite{SymmFunc}. Using the product rule for Schur polynomials, (\ref{eq:2.10}) may be written in terms of the Littlewood-Richardson coefficients 

\begin{eqnarray}
\label{eq:2.11}
	e^{\sum^{\infty}_{m=1}\left(\frac{t^{m}_{1}+t^{m}_{2}}{m}\right)\chi_{\mrn{adj}}(O^{m})} = \sum\limits_{\xi} \sum\limits_{\mu}\sum\limits_{\lambda} (t_{1})^{|2\lambda|/2}(t_{2})^{|2\mu|/2}g(2\lambda,2\mu,\xi) s_{\xi}(y_{1},\cdots , y_{N})
\end{eqnarray}
\\
The partition function (\ref{eq:2.1}) becomes 

\begin{eqnarray}
\label{eq:2.12}
	G_{Sp(N)}(t_{1},t_{2}) \!\!\! &=&\!\!\! \sum\limits_{\xi} \sum\limits_{\mu}\sum\limits_{\lambda} (t_{1})^{|2\lambda|/2}(t_{2})^{|2\mu|/2}g(\lambda,\mu,\xi) \\
\label{eq:2.13}
	&&\!\!\!\!\!\!\times\; \frac{(-1)^{n}}{2^{n}n!}\int_{T_{n}} \prod\limits^{n}_{j=1}\frac{dx_{j}}{2\pi x_{j}} \prod\limits^{n}_{j=1}(x_{j}-x^{-1}_{j})^{2}\Delta(\mrn{x}+\mrn{x}^{-1})^{2}s_{\xi}(x_{1},x^{-1}_{1},\cdots,x_{n},x^{-1}_{n})\nonumber
\end{eqnarray}
\\
The integral in (\ref{eq:2.12}) is 1 for partitions $\xi$ that have even multiplicity, i.e., an even number of boxes in each column, and 0 otherwise \cite{2010arXiv1011.4734V}, \cite{Vanishing}. Therefore we write $\xi^{2}$ for this partition. As we did for $SO(N)$, we succeeded in writing the partition function for $Sp(N)$ gauge theory in terms of the Littlewood-Richardson coefficients

\begin{equation}
\label{eq:2.14}
	G_{Sp(N)}(t_{1},t_{2}) = \sum\limits_{\xi^{2}} \sum\limits_{\mu}\sum\limits_{\lambda} (t_{1})^{|2\lambda|/2}(t_{2})^{|2\mu|/2}g(2\lambda,2\mu,\xi^{2})
\end{equation}
\\
Only partitions $2\mu$ and $2\lambda$, with an even number of boxes in each row, and $\xi^{2}$, with an even number of boxes in each column, contribute to $G_{Sp(N)}$. The Littlewood-Richardson coefficients $g(2\lambda,2\mu,\xi^{2})$ count the number of restricted Schur polynomials that can be defined for labels $(2\lambda, 2\mu,\xi^{2})$. Thus, the counting of restricted Schur polynomials for this class of Young diagram labels matches the counting of states in the free $Sp(N)$ gauge theory.


\section{$Sp(N)$ restricted Schurs}

\subsection{Constructing the operators}

We now give an explicit construction of restricted Schurs for $Sp(N)$ gauge theory. We continue to consider only $N=2n$. The group $Sp(N)$ is the set of $N\times N$ matrices, $S$, satisfying

\begin{equation}
\label{eq:SpNgroupcond}
	S^{T}JS = J, \hspace{20pt} J = \left(\begin{array}{cc}0 & \mathbb{I}_{N/2} \\-\mathbb{I}_{N/2} & 0\end{array}\right). 
\end{equation}
\\
The matrix fields living in the adjoint representation of the $sp(N)$ algebra satisfy

\begin{equation}
\label{eq:algrbracond}
	Z^{T}J + JZ = 0, \hspace{20pt} \mrn{or} \hspace{20pt} Z^{T} = JZJ. 
\end{equation}
\\
Our $Sp(N)$ restricted Schurs must match the counting found in (\ref{eq:2.14}). Since $(r,s)\alpha$ has an even number of boxes in each row, the irrep $([S]_{r},[S]_{s})\alpha$ of $S_{n}[S_{2}]\times S_{m}[S_{2}]$ may be subduced. To match the counting, we construct our operators using the state $\ket{[S]_{r},[S]_{s},\alpha}$. Furthermore, because $R$ has an even number of boxes in each column, the irrep $[A]$ of $S_{q}[S_{2}]$ may be subduced. This leads us to a natural definition for the $Sp(N)$ restricted characters. Define

\begin{equation}
\label{eq:3.0}
	\mc{O}^{Sp(N)}_{R(r,s)\alpha} = \ket{[A]}\!\!\bra{[S]_{r},[S]_{s},\alpha} =\ket{[A]}\!\!\bra{[S]_{r},[S]_{s},\beta}P_{R\rightarrow (r,s)\beta\alpha}
\end{equation}
\\
The $Sp(N)$ restricted character, $\chi^{Sp(N)}_{R(r,s)\alpha}(\sigma)$, clearly has the property

\begin{equation}
\label{eq:3.-1}
	\chi^{Sp(N)}_{R(r,s)\alpha}(\sigma)\mrn{sgn}(\xi) = \chi^{Sp(N)}_{R(r,s)\alpha}(\eta\sigma\xi)\hspace{20pt} \xi \in S_{q}[S_{2}], \eta \in S_{n}[S_{2}]\times S_{m}[S_{2}].
\end{equation}
\\
This function, and its $SO(N)$ counterpart (\ref{eq:9}), resembles the bi-spherical functions discussed in \cite{Ivanov}, \cite{Diaz}, where one of the hyperoctahedral groups in (\ref{eq:3.-1}) is now a subgroup of the other. We construct the restricted Schurs to be invariant under sending $\sigma \rightarrow \eta\sigma\xi$. Define

\begin{equation}
\label{eq:3.1}
	O^{Sp(N)}_{R(r,s)\alpha}(Z,Y) = \frac{1}{\sqrt{n!m!q!2^{2q}}}\sum\limits_{\sigma \in S_{2q}}\chi^{Sp(N)}_{R(r,s)\alpha}(\sigma)\, J_{I}\sigma^{I}_{J}\big((JZ)^{\otimes 2n}\otimes (JY)^{\otimes 2m})^{J}
\end{equation}
\\
where

\begin{equation}
\label{eq:3.2}
	J_{I} = J_{i_{1}i_{2}}J_{i_{3}i_{4}}\cdots J_{i_{2q-1}i_{2q}}
\end{equation}
\\
The quantity $J_{I}(\sigma)^{I}_{J}\big((JZ)^{\otimes 2n}\otimes (JY)^{\otimes 2m})^{J}$ indeed gives all the possible multitrace operators for $n$ $Z$'s, $m$ $Y$'s and for the permutations of $S_{2q}$. For example, 2 $Z$'s and $2$ Y's, give only 4 possible multi-trace operators. They are

\begin{equation}
\label{eq:3.3}
	\mrn{Tr}(Z^{2}Y^{2}), \hspace{10pt} \mrn{Tr}(ZY\!ZY), \hspace{10pt} \mrn{Tr}(Z^{2}) \mrn{Tr}(Y^{2}) \hspace{10pt} \mrn{and} \hspace{10pt} \mrn{Tr}(ZY)^{2}
\end{equation}
\\
Indeed, $J_{I}(\sigma)^{I}_{J}\big((JZ)^{\otimes 4}\otimes (JY)^{\otimes 4})^{J}$ generate all of these for $\sigma \in S_{8}$. Next, we note that  $J_{I}(\sigma)^{I}_{J}\big((JZ)^{\otimes 2n}\otimes (JY)^{\otimes 2m})^{J}$ has the desired symmetry property when $\sigma \rightarrow \eta\sigma\xi$. Since the $J$'s are anti-symmetric under transposition, $\xi$ acting on $J_{I}$ gives $\mrn{sgn}(\xi)J_{I}$, and since $(JZ)$ and $(JY)$ are symmetric under transposition, $\eta$ leaves the tensor invariant. The state $\ket{[A]}$ is calculated as the eigenvector of

\begin{equation}
\label{eq:3.3a}
	P_{[A]} = \frac{1}{q!2^{q}}\sum\limits_{\xi \in S_{q}[S_{2}]} \mrn{sgn}(\xi) \Gamma_{R}(\xi).
\end{equation} 
\\
with eigenvalue 1. The $\ket{[S]_{r},[S]_{s},\alpha}$ is calculated as the eigenvector of

\begin{equation}
\label{eq:3.4}
	P_{R\rightarrow ([S]_{r},[S]_{s})\alpha} = P_{R\rightarrow (r,s)\alpha\alpha}P_{[S,S]}, \hspace{20pt} P_{[S,S]} = \frac{1}{n!m!}\sum\limits_{\eta \in S_{n}[S_{2}]\times S_{m}[S_{2}]}\Gamma_{R}(\eta).
\end{equation}
\\
with eigenvalue 1.

\subsection{Two-point function}

Now consider evaluating the two-point function of the operators in (\ref{eq:3.1}). First, it is straightforward to show that 

\begin{equation}
\label{eq:3.2.0}
	\big{\langle} \big((JZ)^{\otimes 2n} \otimes (JY)^{\otimes 2m}\big)^{J}\big((JZ)^{\otimes 2n} \otimes (JY)^{\otimes 2m}\big)_{L} \big{\rangle} = \sum\limits_{\eta \in S_{n}[S_{2}]\times S_{m}[S_{2}]} \eta^{J}_{L}
\end{equation}
\\
Using (\ref{eq:3.2.0}), the evaluation of the two-point function is analogous to that of the $SO(N)$ case \cite{Kemp:2014apa}. Following the same steps, we arrive at

\begin{equation}
\label{eq:3.2.1}
	\langle O^{Sp(N)}_{R(r,s)\alpha}\overline{O}^{Sp(N)}_{T(t,u)\beta} \rangle = \delta_{RT}\delta_{rt}\delta_{su}\delta_{\alpha\beta} \frac{(2q)!}{d_{R}} \sum\limits_{\psi\in\mc{B}_{q}}\mrn{Tr}\big( P_{[A]}\Gamma_{R}(\psi) \big) J_{I}J^{M}(\psi)^{I}_{M}
\end{equation}
\\
where $\mc{B}_{q}$ is again the set of representatives of the coset $S_{2q}/S_{q}[S_{2}]$, chosen to be the permutations

\begin{equation}
\label{eq:3.2.2}
	\prod\limits^{q-1}_{j=0}\prod\limits^{2j+1}_{i=1}(i,2j+1)
\end{equation}
\\
For the permutations in (\ref{eq:3.2.2}), $J_{I}J^{M}(\psi)^{I}_{M}$ gives exactly the same result as $\delta_{I}\delta^{J}(\psi)^{I}_{J}$. The identity permutation, for example, gives

\begin{eqnarray}
\label{eq:3.2.2.1}
	J_{i_{1}i_{2}}J_{i_{3}i_{4}}\cdots J_{i_{2q-1}i_{2q}} J^{i_{1}i_{2}}J^{i_{3}i_{4}}\cdots J^{i_{2q-1}i_{2q}} &=& (J^{T}J)^{i_{1}}_{i_{1}}(J^{T}J)^{i_{3}}_{i_{3}}\cdots(J^{T}J)^{i_{2q-1}}_{i_{2q-1}}\\
\label{eq:3.2.2.2a}
	&=& \delta^{i_{1}}_{i_{1}}\delta^{i_{3}}_{i_{3}}\cdots \delta^{i_{2q-1}}_{i_{2q-1}}\\
\label{3.2.2.3}
	&=& N^{q}
\end{eqnarray}
\\
A two-cycle $(i,2j+1)$ returns $N^{q-1}$. For example, consider $q=4$ and $(3,5)$. This gives

\begin{equation}
\label{eq:3.2.2.2}
	(J^{T}J)^{i_{1}}_{i_{1}} (J^{T}JJ^{T}J)^{i_{3}}_{i_{3}}(J^{T}J)^{i_{7}}_{i_{7}} = \delta^{i_{1}}_{i_{1}} \delta^{i_{3}}_{i_{3}}\delta^{i_{7}}_{i_{7}}  = N^{3}
\end{equation}
\\
For $(4,7)$, we get

\begin{equation}
\label{eq:3.2.2.2b}
	(J^{T}J)^{i_{1}}_{i_{1}} (J^{2}J^{2})^{i_{3}}_{i_{3}}(J^{T}J)^{i_{5}}_{i_{5}} = \delta^{i_{1}}_{i_{1}} \delta^{i_{3}}_{i_{3}}\delta^{i_{7}}_{i_{7}}  = N^{3}
\end{equation}
\\
For any $\psi \in \mc{B}_{q}$, $J_{I}J^{M}(\psi)^{I}_{M}$ always consists of products of $(J^{T}J)$ and an even number of $J^{2}$'s. If $p$ is the number of two-cycles in $\psi$, then $J_{I}J^{M}(\psi)^{I}_{M}$ gives $N^{q-p}$. After summing over $\mc{B}_{q}$, equation (\ref{eq:3.2.1}) becomes

\begin{equation}
\label{eq:3.2.32}
	\langle O^{Sp(N)}_{R(r,s)\alpha}\overline{O}^{Sp(N)}_{T(t,u)\beta} \rangle = \delta_{RT}\delta_{rt}\delta_{su}\delta_{\alpha\beta} \frac{d_{R}}{(2q)!} \langle [A] | \prod\limits^{q-1}_{j=0}(N + J_{2j+1}) |[A]\rangle
\end{equation}
\\
Instead of giving a product of weights coming from the odd columns in $R$, as it did for $SO(N)$, (\ref{eq:3.2.32}) now gives the product of weights coming from the odd \emph{rows} of $R$. The two-point function for the $Sp(N)$ restricted Schurs is

\begin{equation}
\label{eq:3.2.3111}
	\langle O^{Sp(N)}_{R(r,s)\alpha}\overline{O}^{Sp(N)}_{T(t,u)\beta} \rangle = \delta_{RT}\delta_{rt}\delta_{su}\delta_{\alpha\beta} \frac{(2q)!}{d_{R}} \prod\limits_{i\;\in \mrn{odd\;rows\;in\;R}}c_{i}
\end{equation}
\\
This gives precisely the same result as the $SO(N)$ case, but with $N \rightarrow -N$, as expected. Thus, to calculate the two-point function of operators $O^{Sp(N)}_{R(r,s)\alpha}$, calculate the two-point function of $O^{SO(N)}_{R_{c}(r_{c},s_{c})\alpha}$ in $SO(N)$ gauge theory, and send $N \rightarrow -N$. This gives the result for $Sp(N)$. By $R_{c}$, we mean the conjugate (or transpose) of the Young diagram. As a check of these conclusions, we present two examples in appendix B.


\section{Exact multi-trace correlators with two scalar fields}

In this section, we express any multi-trace operator involving $Z$'s and $Y$'s as a linear combination of the $SO(N)$ and $Sp(N)$ restricted Schur polynomials, (\ref{eq:2}) and (\ref{eq:3.1}). Thereafter, we derive a product rule for these operators. First, let's discuss $SO(N)$.
 
Define the `dual restricted character' 

\begin{equation}
\label{eq:1}
	\chi^{R(r,s)\alpha}_{SO(N)}(\sigma) \equiv \frac{d_{R}\sqrt{n!m!q!2^{2q}}}{(2q)!}\mrn{Tr}\big( \mc{O}^{T}_{R(r,s)\alpha}\Gamma_{R}(\sigma^{-1}) \big) 
\end{equation}
\\
The aim is to use (\ref{eq:1}) to express any multi-trace operator in terms of the restricted Schurs in (\ref{eq:2}). The calculation is analogous to that in \cite{Bhattacharyya:2008xy}. Calculate

\begin{eqnarray}
\label{eq:3}
	\sum\limits_{R,r,s,\alpha}\chi^{R(r,s)\alpha}_{SO(N)}(\sigma)\mrn{Tr}\big( \mc{O}_{R(r,s)\alpha}\Gamma_{R}(\tau) \big) = \frac{\sqrt{n!m!q!2^{2q}}}{(2q)!} \sum\limits_{R,r,s,\alpha} d_{R} \mrn{Tr}\big( \mc{O}^{T}_{R(r,s)\alpha}\Gamma_{R}(\sigma^{-1})\big)\mrn{Tr}\big( \mc{O}_{R(r,s)\alpha}\Gamma_{R}(\tau) \big)
\end{eqnarray}
\\
The $\Gamma$'s may be expanded in the $R$ basis in the following way

\begin{equation}
\label{eq:4}
	\Gamma_{R}(\sigma) = \sum\limits_{I,J} \ket{R,I}\!\!\bra{R,J}\, \Gamma_{R}(\sigma)_{IJ}
\end{equation}
\\
Using this, the trace $\mrn{Tr}\big( \mc{O}_{R(r,s)\alpha}\Gamma_{R}(\tau) \big)$ may be written as

\begin{eqnarray}
	\label{eq:5}
	\mrn{Tr}\big( \mc{O}_{R(r,s)\alpha}\Gamma_{R}(\tau) \big) &=& \sum\limits_{K}\bra{R,K}\mc{O}_{R(r,s)\alpha}\Gamma_{R}(\tau) \ket{R,K}\\
	\label{eq:6}
	&=&  \sum\limits_{K,L}\bra{R,K}\mc{O}_{R(r,s)\alpha} \ket{R,L} \Gamma_{R}(\tau)_{LK}
\end{eqnarray}
\\
Then

\begin{eqnarray}
	\label{eq:7}
	\sum\limits_{r,s,\alpha} \mrn{Tr}\big( \mc{O}^{T}_{R(r,s)\alpha}\Gamma_{R}(\sigma^{-1})\big)\mrn{Tr}\big( \mc{O}_{R(r,s)\alpha}\Gamma_{R}(\tau) \big) &=& \sum\limits_{r,s,\alpha} \Big( \sum\limits_{I,J}\bra{R,I}\mc{O}^{T}_{R(r,s)\alpha} \ket{R,J} \Gamma_{R}(\sigma^{-1})_{JI}\Big)\nonumber \\
	\label{eq:8}
	&&\times \Big( \sum\limits_{K,L}\bra{R,K}\mc{O}_{R(r,s)\alpha} \ket{R,L} \Gamma_{R}(\tau)_{LK}\Big)\nonumber\\
	\label{eq:11}
	&=&  \sum\limits_{r,s,\alpha} \sum\limits_{I,J,K,L} \Gamma_{R}(\sigma^{-1})_{JI}\Gamma_{R}(\tau)_{LK}\nonumber \\
	&&\!\!\!\! \times \langle R,I | [A]_{r},[A]_{s},\alpha \rangle\langle [S] |R,J \rangle \langle R,K | [S]\rangle\langle [A]_{r},[A]_{s},\alpha  |R,L \rangle  \nonumber \\
	\label{eq:12}
	&=& \sum\limits_{r,s,\alpha} \sum\limits_{I,J,K,L} \Gamma_{R}(\sigma^{-1})_{JI}\Gamma_{R}(\tau)_{LK}\nonumber \\
	&&\!\!\!\!\times \langle R,K | [S] \rangle\!\langle [S] |R,J \rangle\; \langle R,I | [A]_{r},[A]_{s},\alpha\rangle\!\langle [A]_{r},[A]_{s},\alpha | R,L \rangle \nonumber\\
	\label{eq:18a}
\end{eqnarray}
\\
Recognising

\begin{eqnarray}
\label{eq:13}
	P_{[S]} &=& \frac{1}{q!2^{q}}\sum\limits_{\xi\in S_{q}[S_{2}]}\Gamma_{R}(\xi) = \ket{[S]}\!\!\bra{[S]} \\
\label{eq:14}
	P_{[A,A]} &=& \frac{1}{n!m!2^{q}}\sum\limits_{\eta\in S_{n}[S_{2}]\times S_{m}[S_{2}]}\mrn{sgn}(\eta)\Gamma_{R}(\eta)\\
\label{eq:15}
	&=& \sum\limits_{r,s,\alpha} \ket {[A]_{r},[A]_{s},\alpha}\!\!\bra {[A]_{r},[A]_{s},\alpha} 
\end{eqnarray}
\\
In appendix A, we discuss going from (\ref{eq:14}) to (\ref{eq:15}) in more detail. Equation (\ref{eq:18a}) then becomes 

\begin{eqnarray}
	\label{eq:16}
	\sum\limits_{r,s,\alpha} \mrn{Tr}\big( \mc{O}^{T}_{R(r,s)\alpha}\Gamma_{R}(\sigma^{-1})\big)\mrn{Tr}\big( \mc{O}_{R(r,s)\alpha}\Gamma_{R}(\tau) \big) \!\!\!\!&=&\!\!\!\! \frac{1}{n!m!q!2^{2q}}\sum\limits_{\xi\in S_{q}[S_{2}]}\sum\limits_{\eta\in S_{n[S_{2}]\times S_{m}[S_{2}]}}\sum\limits_{I,J,K,L}\nonumber\\
	&&\times \mrn{sgn}(\eta) \Gamma_{R}(\sigma^{-1})_{JI}\Gamma_{R}(\tau)_{LK}\Gamma_{R}(\xi)_{KJ}\Gamma_{R}(\eta)_{IL} \nonumber\\
\label{eq:17}
	&=&  \frac{1}{n!m!q!2^{2q}} \sum\limits_{\xi\in S_{q}[S_{2}]}\sum\limits_{\eta\in S_{n[S_{2}]\times S_{m}[S_{2}]}} \mrn{sgn}(\eta) \chi_{R}(\sigma^{-1}\eta\tau\xi)\nonumber\\
\label{eq:18}
\end{eqnarray}
\\
Using (\ref{eq:18}), equation (\ref{eq:3}) becomes

\begin{eqnarray}
\label{eq:19}
	\sum\limits_{R,r,s,\alpha}\chi^{R(r,s)\alpha}_{SO(N)}(\sigma)\mrn{Tr}\big( \mc{O}_{R(r,s)\alpha}\Gamma_{R}(\tau) \big) = \frac{1}{\sqrt{n!m!q!2^{2q}}} \sum\limits_{R} \frac{d_{R}}{(2q)!} \sum\limits_{\xi\in S_{q}[S_{2}]}\sum\limits_{\eta\in S_{n[S_{2}]\times S_{m}[S_{2}]}} \mrn{sgn}(\eta) \chi_{R}(\sigma^{-1}\eta\tau\xi)
\end{eqnarray}
\\
In this expression, we are summing over all possible $R \vdash 2q$ and terms for which $R$ does not have an even number of boxes in each row vanish\footnote{This is a simple consequence of the fact that only $R$ having even rows is capable of subducing the $S_{q}[S_{2}]$ irrep $[S]$.}.

\begin{eqnarray}
\label{eq:20}
	\sum\limits_{R,r,s,\alpha}\chi^{R(r,s)\alpha}_{SO(N)}(\sigma)\mrn{Tr}\big( \mc{O}_{R(r,s)\alpha}\Gamma_{R}(\tau) \big) \!\!\!\! &=& \!\!\!\!\frac{1}{\sqrt{n!m!q!2^{2q}}}  \sum\limits_{\xi\in S_{q}[S_{2}]}\sum\limits_{\eta\in S_{n[S_{2}]\times S_{m}[S_{2}]}} \mrn{sgn}(\eta)  \sum\limits_{R} \frac{d_{R}}{(2q)!}  \chi_{R}(\sigma^{-1}\eta\tau\xi) \nonumber \\
\label{eq:21}
	\!\!\!\! &=& \!\!\!\! \frac{1}{\sqrt{n!m!q!2^{2q}}} \sum\limits_{\xi\in S_{q}[S_{2}]}\sum\limits_{\eta\in S_{n[S_{2}]\times S_{m}[S_{2}]}} \mrn{sgn}(\eta) \delta(\sigma^{-1}\eta\tau\xi)
\end{eqnarray}
\\
We now use result (\ref{eq:21}) to write any multi-trace operator in terms of $O_{R(r,s)\alpha}(Z,Y)$. Consider

\begin{eqnarray}
\label{eq:22}
	\sum\limits_{R(r,s)\alpha} \chi^{R(r,s)\alpha}_{SO(N)}(\sigma) O_{R(r,s)\alpha}(Z,Y) &=& \frac{1}{\sqrt{n!m!q!2^{2q}}}\sum\limits_{\tau\in S_{2q}}\bigg[ \sum\limits_{R(r,s)\alpha}\chi^{R(r,s)\alpha}_{SO(N)}(\sigma)\mrn{Tr}\big( \mc{O}_{R(r,s)\alpha}\Gamma_{R}(\tau) \big) \bigg]\nonumber\\
\label{eq:23}
	&& \hspace{100pt} \times \; C^{4\nu}_{I}\tau^{I}_{J}(Z^{\otimes 2n}\otimes Y^{\otimes 2m})^{J}\nonumber\\
\label{eq:24a}
	&=& \frac{1}{n!m!q!2^{2q}}\sum\limits_{\tau\in S_{2q}} \sum\limits_{\xi\in S_{q}[S_{2}]}\sum\limits_{\eta\in S_{n[S_{2}]\times S_{m}[S_{2}]}} \mrn{sgn}(\eta) \delta(\sigma^{-1}\eta\tau\xi)\nonumber\\
\label{eq:25a}
	&& \hspace{100pt} \times \;C^{4\nu}_{I}\tau^{I}_{J}(Z^{\otimes 2n}\otimes Y^{\otimes 2m})^{J}
\end{eqnarray}
\\
Summing over $\tau$, $\delta(\sigma^{-1}\eta\tau\xi)$ sets $\tau = \eta^{-1}\sigma\xi^{-1}$. Thus

\begin{eqnarray}
\label{eq:2211}
	\sum\limits_{R(r,s)\alpha} \chi^{R(r,s)\alpha}_{SO(N)}(\sigma) O_{R(r,s)\alpha}(Z,Y) \!\!\!\! &=&\!\!\!\! \frac{1}{n!m!q!2^{2q}} \sum\limits_{\xi\in S_{q}[S_{2}]}\sum\limits_{\eta\in S_{n[S_{2}]\times S_{m}[S_{2}]}} \mrn{sgn}(\eta) C^{4\nu}_{I}(\eta^{-1}\sigma\xi^{-1})^{I}_{J}(Z^{\otimes 2n}\otimes Y^{\otimes 2m})^{J}\nonumber\\
	\label{eq:2311}
\end{eqnarray}
\\
The tensor $C^{4\nu}_{I}$ acted on by the permutation $\xi \in S_{q}[S_{2}]$ is invariant, and the $\eta \in S_{n}[S_{2}]\times S_{m}[S_{2}]$ acting on the $ZY$ tensor picks up a $\mrn{sgn}(\eta)$. After summing over $\xi$ and $\eta$, we obtain

\begin{equation}
\label{eq:26}
	\sum\limits_{R(r,s)\alpha} \chi^{R(r,s)\alpha}_{SO(N)}(\sigma) O_{R(r,s)\alpha}(Z,Y) =  C^{4\nu}_{K} \sigma^{K}_{L}(Z^{\otimes 2n}\otimes Y^{\otimes 2m})^{L}
\end{equation}
\\
Each $\sigma$ gives some multi-trace operator. Using the restricted Schurs in (\ref{eq:2}) for the $S_{8}$ irreps,

\begin{eqnarray}
\label{eq:26.1}
	&&\ytableausetup{boxsize=0.8em}\ydiagram[]{4,4}, \Big(\,\ytableausetup{boxsize=0.8em}\ydiagram[]{2,2}\,,\,\ytableausetup{boxsize=0.8em}\ydiagram[]{2,2}\,\Big) \hspace{20pt} \ytableausetup{boxsize=0.8em}\ydiagram[]{2,2,2,2}, \Big(\,\ytableausetup{boxsize=0.8em}\ydiagram[]{2,2}\,,\,\ytableausetup{boxsize=0.8em}\ydiagram[]{2,2}\,\Big) \hspace{20pt} \ytableausetup{boxsize=0.8em}\ydiagram[]{2,2,2,2}, \Big(\,\ytableausetup{boxsize=0.8em}\ydiagram[]{1,1,1,1}\,,\,\ytableausetup{boxsize=0.8em}\ydiagram[]{1,1,1,1}\,\Big)\\
\label{eq:26.2}
	&& \ytableausetup{boxsize=0.8em}\ydiagram[]{4,2,2}, \Big(\,\ytableausetup{boxsize=0.8em}\ydiagram[]{2,2}\,,\,\ytableausetup{boxsize=0.8em}\ydiagram[]{2,2}\,\Big)\nonumber
\end{eqnarray}
\\
we have checked formula (\ref{eq:26}) for a large number of permutations $\sigma$. We present two examples in appendix C.

The calculation is exactly the same for $Sp(N)$. Defining the $Sp(N)$ `dual restricted character'

\begin{equation}
\label{eq:27a}
	\chi^{R(r,s)\alpha}_{Sp(N)}(\sigma) = \frac{d_{R}\sqrt{n!m!q!2^{2q}}}{(2q)!}\mrn{Tr}\big( \mc{O}^{Sp(N)\,T}_{R(r,s)}\Gamma_{R}(\sigma^{-1}) \big),
\end{equation}
\\
we find that any multi-trace operator may be written as

\begin{equation}
\label{eq:28}
	\sum\limits_{R,r,s,\alpha} \chi^{R(r,s)\alpha}_{Sp(N)}(\sigma) O^{Sp(N)}_{R(r,s)\alpha}(Z,Y) = J_{I}(\sigma)^{I}_{K}\big( (JZ)^{\otimes 2n} \otimes (JY)^{\otimes 2m} \big)^{K}
\end{equation}
\\
We are now able to express any multi-trace operator, involving two scalar fields, in the free $SO(N)$ and $Sp(N)$ theory in terms of our restricted Schur basis.

A product rule for operators $O^{SO(N)}_{R(r,s)\alpha}(Z,Y)$ in terms of restricted Littlewood-Richardson coefficients is easily derived. The basic idea is exactly the same as in \cite{Bhattacharyya:2008xy}. We multiply two restricted Schurs, one having labels $R_{1}(r_{1},s_{1})\alpha_{1}$ and the other $R_{2}(r_{2},s_{2})\alpha_{2}$, to produce a linear combination of restricted Schurs with labels $R_{1}+R_{2}, (r_{1}+r_{2},s_{1}+s_{2})\beta$. $R_{1}$ and $R_{2}$ are irreps of $S_{2q_{1}}$ and $S_{2q_{2}}$ respectively. $(r_{1},s_{1})$ and $(r_{2},s_{2})$ are irreps of $S_{2n_{1}}\times S_{2m_{1}}$ and $S_{2n_{2}}\times S_{2m_{2}}$ respectively, where $q_{i}=n_{i}+m_{i}$. $\alpha_{1}$ and $\alpha_{2}$ are the multiplicity labels for the two respective subgroup irreps. Thus, $R_{i}(r_{i},s_{i})\alpha_{i}$ defines a restricted Schur having $n_{i}$ $Z$'s and $m_{i}$ $Y$'s. $R_{1}+R_{2}$ is an irrep of $S_{2q_{1}+2q_{2}}$, $r_{1}+r_{2}$ is an irrep of $S_{2n_{1}+2n_{2}}$ and $s_{1}+s_{2}$ is an irrep of $S_{2m_{1}+2m_{2}}$. $\beta$ simply labels the $(r_{1}+r_{2},s_{1}+s_{2})$ copy subduced from $R_{1}+R_{2}$. The set of labels $\{R_{1}+R_{2},(r_{1}+r_{2},s_{1}+s_{2}),\beta\}$ defines a restricted Schur having $n_{1}+n_{2}$ $Z$'s and $m_{1}+m_{2}$ $Y$'s. As in \cite{Bhattacharyya:2008xy}, it is convenient to streamline the notation. Thus, denote $\{i\} \equiv R_{i}(r_{i},s_{i})\alpha_{i}$ and $\{1+2\} \equiv (R_{1}+R_{2})(r_{1}+r_{2},s_{1}+s_{2})\beta$. Also, write $n_{12} = n_{1} + n_{2}$, $m_{12} = m_{1}+ m_{2}$ and $q_{12} = q_{1}+q_{2}$. Thus, for example, (\ref{eq:Newdefintermsofdelta}) may be written as

\begin{equation}
\label{eq:BigRS}
	O^{SO(N)}_{\{1+2\}}(Z,Y) = \frac{1}{\sqrt{n_{12}!m_{12}!q_{12}!2^{2q_{12}}}}\sum\limits_{\rho \in S_{2q_{12}}} \chi^{SO(N)}_{\{1+2\}}(\rho) \delta_{I}\rho^{I}_{J} \big( Z^{\otimes 2n_{12}} \otimes Y^{\otimes 2m_{12}} \big)^{J}
\end{equation}
\\
In the following derivation, we use the operators defined in (\ref{eq:Newdefintermsofdelta}) which are equivalent to the ones defined in (\ref{eq:2}). The factorisation we need in our product rule occurs naturally for the operators defined using the $\delta_{I}$ tensor, rather than the $C^{4\nu}_{I}$ tensor. The following derivation is for $SO(N)$, but the derivation for $Sp(N)$ is the same. Define the restricted Littlewood-Richardson coefficients to be

\begin{equation}
\label{eq:reslittwood}
	f^{\{1+2\}}_{\{1\}, \{2\}} = \frac{1}{\sqrt{n_{1}!m_{1}!n_{2}!m_{1}!m_{2}!2^{2q_{1}+2q_2}}}\sum\limits_{\sigma_{1}\in S_{2q_{1}}}\sum\limits_{\sigma_{2}\in S_{2q_{2}}}\chi^{SO(N)}_{\{1\}}(\sigma_{1})\chi^{SO(N)}_{\{2\}}(\sigma_{2})\chi^{\{1+2\}}_{SO(N)}(\sigma_{1}\cdot \sigma_{2})
\end{equation}
\\
where $\chi^{\{1+2\}}_{SO(N)}$ is the dual restricted character defined in (\ref{eq:1}). We want to evaluate 

\begin{equation}
\label{eq:Frodo}
	\sum\limits_{\{1+2\}}f^{\{1+2\}}_{\{1\},\{2\}}O^{SO(N)}_{\{1+2\}}(Z,Y).
\end{equation}
\\
To evaluate (\ref{eq:Frodo}), we use equation (\ref{eq:21}) to write

\begin{eqnarray}
\label{eq:1001}
	\sum\limits_{\{1+2\}}\chi^{\{1+2\}}_{SO(N)}(\sigma_{1}\cdot \sigma_{2})\chi^{SO(N)}_{\{1+2\}}(\rho) &=& \frac{1}{\sqrt{n_{12}!m_{12}!q_{12}!2^{q_{12}!}}}\sum\limits_{\xi \in S_{q_{12}}[S_{2}]}\sum\limits_{\eta \in S_{n_{12}}[S_{2}]\times S_{m_{12}}[S_{2}]}\mrn{sgn}(\eta) \nonumber \\
	&& \hspace{100pt} \times \; \delta(\sigma^{-1}_{1}\!\cdot\! \sigma^{-1}_{2}\eta\rho\xi)
\end{eqnarray}
\\
Summing over $\rho$ sets $\rho = \eta^{-1}\sigma_{1}\sigma_{2}\xi^{-1}$. As before, $\xi$ acting on $\delta_{I}$ is invariant and $\eta$ acting on the $ZY$ tensor gives back an extra $\mrn{sgn}(\eta)$. Summing over $\xi$ and $\eta$ then cancels the normalisation factor in (\ref{eq:BigRS}), and thus we write

\begin{eqnarray}
\label{eq:1002}
	\sum\limits_{\{1+2\}}f^{\{1+2\}}_{\{1\},\{2\}}O^{SO(N)}_{\{1+2\}}(Z,Y) \!\!\!\! &=& \!\!\!\! \frac{1}{\sqrt{n_{1}!m_{1}!n_{2}!m_{1}!m_{2}!2^{2q_{1}+2q_2}}}\sum\limits_{\sigma_{1}\in S_{2q_{1}}}\sum\limits_{\sigma_{2}\in S_{2q_{2}}}\chi^{SO(N)}_{\{1\}}(\sigma_{1})\chi^{SO(N)}_{\{2\}}(\sigma_{2}) \nonumber \\
\label{eq:1003}
	&& \delta_{I}(\sigma_{1}\cdot \sigma_{2})^{I}_{J}\big( Z^{\otimes 2n_{12}}\otimes Y^{\otimes 2m_{12}} \big)^{J}
\end{eqnarray}
\\
By $\sigma_{1}$, we mean the permutation that acts on the first $2n_{1}$ $Z$ indices and first $2m_{1}$ $Y$ indices. By $\sigma_{2}$, we mean the permutation that acts on the second $2n_{2}$ $Z$ indices and second $2m_{2}$ $Y$ indices. The second line in (\ref{eq:1003}) factories and we may write

\begin{eqnarray}
\label{eq:1004}
	\sum\limits_{\{1+2\}}f^{\{1+2\}}_{\{1\},\{2\}}O^{SO(N)}_{\{1+2\}}(Z,Y) \!\!\!\! &=&\!\!\!\! \frac{1}{\sqrt{n_{1}!m_{1}!q_{1}!2^{2q_{1}}}}\sum\limits_{\sigma_{1}\in S_{2q_{1}}}\chi^{SO(N)}_{\{1\}}(\sigma_{1})\delta_{I_{1}} (\sigma_{1})^{I_{1}}_{J_{1}}\big(Z^{\otimes 2n_{1}} \otimes Y^{\otimes 2m_{1}} \big)^{J_{1}}\nonumber \\
	\label{eq:1005}
	&&\!\!\!\!\times \frac{1}{\sqrt{n_{2}!m_{2}!q_{2}!2^{2q_{2}}}}\sum\limits_{\sigma_{2}\in S_{2q_{2}}}\chi^{SO(N)}_{\{2\}}(\sigma_{2})\delta_{I_{2}} (\sigma_{2})^{I_{2}}_{J_{2}}\big(Z^{\otimes 2n_{2}} \otimes Y^{\otimes 2m_{2}} \big)^{J_{2}}\nonumber \\
	\label{eq:1006}
\end{eqnarray}
\\
Thus, we have achieved the desired result

\begin{equation}
\label{eq:productrule}
	\sum\limits_{\{1+2\}}f^{\{1+2\}}_{\{1\},\{2\}}O^{SO(N)}_{\{1+2\}}(Z,Y) = O^{SO(N)}_{\{1\}}(Z,Y)O^{SO(N)}_{\{2\}}(Z,Y).
\end{equation}
\\
In appendix D, we check this product with a simple example.


\section{Discussion}

In this work, we have defined a basis for the $1/4$-BPS sector of the free super Yang-Mills theory with a symplectic gauge group. These operators are very similar to those defined for the theory with orthogonal gauge group. The difference between the two cases is that the symmetrisations of the irreducible representations defining the operators have been exchanged, as expected. The two-point function for the symplectic gauge theory operators was related to its orthogonal gauge theory counterpart by replacing $N$ by $-N$.

The results of this work make it possible to compute correlation functions of any kind of multi-matrix, multi-trace operators involving two scalar fields. In such a correlation function, each trace operator may be expressed in terms of restricted Schur polynomials. Using the product rule derived above, computing the correlation function of many restricted Schurs may be transformed into a simple two-point function computation, the formula for which is given in (\ref{eq:27}) and (\ref{eq:3.2.3111}).

Studying the spectrum of anomalous dimensions of our restricted Schurs is an interesting problem, especially in the limit that these operators become dual to excited giant gravitons. Such a study may yield new insights into the non-perturbative physics of their D-brane duals. Pursuing this direction may also allow us to make some concrete statements about whether or not integrability is preserved in non-planar limits of the $SO(N)$ or $Sp(N)$ gauge theory.\\

\emph{Acknowledgments:} I would like to thank Prof. Robert de Mello Koch whose guidance and support has made this work possible.


\appendix

\section{An expression for $P_{[A,A]}$}

In this section, we try prove that

\begin{equation}
\label{eq:A11}
	P_{[A,A]} = \frac{1}{n!m!2^{q}} \sum\limits_{\eta \in S_{n}[S_{2}]\times S_{m}[S_{2}]}\mrn{sgn}(\eta)\Gamma_{R}(\eta) = \sum\limits_{r,s,\nu} \ket{[A]_{r},[A]_{s},\nu}\!\!\bra{[A]_{r},[A]_{s},\nu}
\end{equation}
\\
Since $\eta_{1}\cdot \eta_{2} \in S_{2n}\times S_{2m}$, we can write \cite{Quiver}

\begin{eqnarray}
\label{eq:A12}
	\Gamma_{R}(\eta_{1}\cdot \eta_{2})_{IJ} = \sum\limits_{r,s,\nu}\sum\limits_{j_{1}j_{2}k_{1}k_{2}}\langle R,I | R(r,s)\nu;j_{1},j_{2} \,\rangle \Gamma_{r}(\eta_{1})_{j_{1}k_{1}}\Gamma_{s}(\eta_{2})_{j_{2}k_{2}} \langle R(r,s)\nu;k_{1},k_{2} | R,J \rangle\nonumber\\
\label{eq:A13}
\end{eqnarray}
\\
Writing out the $IJ$-th matrix element of $P_{[A,A]}$,

\begin{equation}
\label{eq:A14}
	(P_{[A,A]})_{IJ} = \bra{R,I} \bigg( \sum\limits_{r,s,\nu}(P_{[A]_{r}})_{j_{1}k_{1}}(P_{[A]_{s}})_{j_{2}k_{2}}\ket{R(r,s)\nu;j_{1},j_{2}}\!\!\bra{R(r,s)\nu;k_{1},k_{2}}\bigg) \ket{R,J}
\end{equation}
\\
where we wrote

\begin{equation}
\label{eq:A14a}
	P_{[A]_{r}} = \frac{1}{n!2^{n}}\sum\limits_{\eta_{1}\in S_{n}[S_{2}]}\mrn{sgn}(\eta_{1})\Gamma_{r}(\eta_{1})
\end{equation}
\\
with a similar expression for $P_{[A]_{s}}$. Using the fact the $P$'s are projectors, i.e., $P^{2}_{[A]} = P_{[A]}$, this matrix element becomes

\begin{eqnarray}
\label{eq:A15}
	(P_{[A,A]})_{IJ} = \bra{R,I} \bigg( \!\sum\limits_{r,s,\nu}(P_{[A]_{r}})_{j_{1}l_{1}}(P_{[A]_{r}})_{l_{1}k_{1}}(P_{[A]_{s}})_{j_{2}l_{2}}(P_{[A]_{s}})_{l_{2}k_{2}}\ket{R(r,s)\nu;j_{1},j_{2}}\!\!\bra{R(r,s)\nu;k_{1},k_{2}}\!\bigg)\! \ket{R,J}\nonumber\\
\label{eq:A16}
\end{eqnarray}
\\
We then have

\begin{eqnarray}
\label{eq:A17}
	(P_{[A,A]})_{IJ} &=&\!\! \bra{R,I} \bigg( \sum\limits_{r,s,\nu} \sum\limits_{l_{1}l_{2}}\ket{[A]_{r},[A]_{s},\nu;l_{1}l_{2}}\!\!\bra{[A]_{r},[A]_{s},\nu};l_{1}l_{2} \bigg)\! \ket{R,J}\nonumber\\
\label{eq:A18}
	&=& \sum\limits_{r,s,\nu} \langle R,I | [A]_{r},[A]_{s},\nu\rangle \! \langle [A]_{r},[A]_{s},\nu | R,J\rangle
\end{eqnarray}
\\
To get the last line, we used the fact that $([A],[A])$ is a 1-dimensional irrep. 


\section{$Sp(N)$ two-function examples}

First, consider for $Sp(N)$

\begin{equation}
\label{eq:ex1}
	R = \ytableausetup{boxsize=1em}\ydiagram[]{2,2}, (r,s) = \big(\, \ytableausetup{boxsize=1em}\ydiagram[]{2}\,,\, \ytableausetup{boxsize=1em}\ydiagram[]{2} \,\big)
\end{equation}
\\
The restricted Schur (\ref{eq:3.1}) for this operator was calculated to be

\begin{equation}
\label{eq:ex2}
	O^{Sp(N)}_{\ytableausetup{boxsize=0.30em}\ydiagram[]{2,2}\,(\,\ytableausetup{boxsize=0.30em}\ydiagram[]{2}\,,\,\ytableausetup{boxsize=0.30em}\ydiagram[]{2}\,)} = -\sqrt{6} \mrn{Tr}(ZY)
\end{equation}
 \\
 It's two-point function is 
 
 \begin{equation}
\label{eq:ex3}
	\langle O^{Sp(N)}_{\ytableausetup{boxsize=0.30em}\ydiagram[]{2,2}\,(\,\ytableausetup{boxsize=0.30em}\ydiagram[]{2}\,,\,\ytableausetup{boxsize=0.30em}\ydiagram[]{2}\,)} \overline{O}^{Sp(N)}_{\ytableausetup{boxsize=0.30em}\ydiagram[]{2,2}\,(\,\ytableausetup{boxsize=0.30em}\ydiagram[]{2}\,,\,\ytableausetup{boxsize=0.30em}\ydiagram[]{2}\,)} \rangle = 12N(N+1).
\end{equation}
\\
Firstly, this agrees with (\ref{eq:3.2.3111}) and secondly, we compare this to the $SO(N)$ two-point function for 

\begin{equation}
\label{eq:ex4}
	R_{c} = \ytableausetup{boxsize=1em}\ydiagram[]{2,2}, (r_{c},s_{c}) = \Big(\, \ytableausetup{boxsize=0.9em}\ydiagram[]{1,1}\,,\, \ytableausetup{boxsize=0.9em}\ydiagram[]{1,1}\, \Big)
\end{equation}
 \\
The restricted Schur for $SO(N)$ (\ref{eq:2}) is (see \cite{Kemp:2014apa})

\begin{equation}
\label{ex5}
	O^{SO(N)}_{\ytableausetup{boxsize=0.30em}\ydiagram[]{2,2}\,(\,\ytableausetup{boxsize=0.30em}\ydiagram[]{1,1}\,,\,\ytableausetup{boxsize=0.30em}\ydiagram[]{1,1}\,)} = \sqrt{6}\mrn{Tr}(ZY)
\end{equation}
\\
with two-point function

\begin{equation}
\label{eq:ex6}
	\langle O^{SO(N)}_{\ytableausetup{boxsize=0.30em}\ydiagram[]{2,2}\,(\,\ytableausetup{boxsize=0.30em}\ydiagram[]{1,1}\,,\,\ytableausetup{boxsize=0.30em}\ydiagram[]{1,1}\,)} \overline{O}^{SO(N)}_{\ytableausetup{boxsize=0.30em}\ydiagram[]{2,2}\,(\,\ytableausetup{boxsize=0.30em}\ydiagram[]{1,1}\,,\,\ytableausetup{boxsize=0.30em}\ydiagram[]{1,1}\,)} \rangle = 12N(N-1).
\end{equation}
\\
Clearly, sending $N$ to $-N$ gives us the $Sp(N)$ result (\ref{eq:ex3}). 

Next, consider for $Sp(N)$

\begin{equation}
\label{eq:ex7}
	R = \ytableausetup{boxsize=1em}\ydiagram[]{4,4}, (r,s) = \big(\, \ytableausetup{boxsize=1em}\ydiagram[]{4}\,,\, \ytableausetup{boxsize=1em}\ydiagram[]{4} \,\big)
\end{equation}
\\
The restricted Schur was calculated to be

\begin{equation}
\label{eq:ex8}
	O^{Sp(N)}_{\ytableausetup{boxsize=0.30em}\ydiagram[]{4,4}\,(\ytableausetup{boxsize=0.30em}\ydiagram[]{4}\,,\,\ytableausetup{boxsize=0.30em}\ydiagram[]{4}\,)} = 2\sqrt{30} \Big(\mrn{Tr}(ZY)^{2} + 2\mrn{Tr}(ZY\!ZY)\Big)
\end{equation}
 \\
 Its two-point function is 
 
\begin{equation}
\label{eq:ex9}
	\langle O^{Sp(N)}_{\ytableausetup{boxsize=0.30em}\ydiagram[]{4,4}\,(\ytableausetup{boxsize=0.30em}\ydiagram[]{4}\,,\,\ytableausetup{boxsize=0.30em}\ydiagram[]{4}\,)} \overline{O}^{Sp(N)}_{\ytableausetup{boxsize=0.30em}\ydiagram[]{4,4}\,(\ytableausetup{boxsize=0.30em}\ydiagram[]{4}\,,\,\ytableausetup{boxsize=0.30em}\ydiagram[]{4}\,)} \rangle = 2880N(N+1)(N+2)(N+3),
\end{equation}
\\
agreeing with (\ref{eq:3.2.3111}). We now compare this with the $SO(N)$ two-point function for 

\begin{equation}
\label{eq:ex4}
	R_{c} = \ytableausetup{boxsize=1em}\ydiagram[]{2,2,2,2}, (r_{c},s_{c}) = \Big(\, \ytableausetup{boxsize=0.9em}\ydiagram[]{1,1,1,1}\,,\, \ytableausetup{boxsize=0.9em}\ydiagram[]{1,1,1,1}\, \Big)
\end{equation}
 \\
The restricted Schur (\ref{eq:2}) is

\begin{equation}
\label{ex5}
	O^{SO(N)}_{ \ytableausetup{boxsize=0.3em}\ydiagram[]{2,2,2,2}\,(\,\ytableausetup{boxsize=0.3em}\ydiagram[]{1,1,1,1}\,,\ytableausetup{boxsize=0.3em}\ydiagram[]{1,1,1,1}\,)} = 2\sqrt{30}\Big( \mrn{Tr}(ZY)^{2} - 2\mrn{Tr}(ZY\!ZY)\Big)
\end{equation}
\\
with two-point function

\begin{equation}
\label{eq:ex6}
	\langle O^{SO(N)}_{ \ytableausetup{boxsize=0.3em}\ydiagram[]{2,2,2,2}\,(\,\ytableausetup{boxsize=0.3em}\ydiagram[]{1,1,1,1}\,,\ytableausetup{boxsize=0.3em}\ydiagram[]{1,1,1,1}\,)} \overline{O}^{SO(N)}_{ \ytableausetup{boxsize=0.3em}\ydiagram[]{2,2,2,2}\,(\,\ytableausetup{boxsize=0.3em}\ydiagram[]{1,1,1,1}\,,\ytableausetup{boxsize=0.3em}\ydiagram[]{1,1,1,1}\,)} \rangle = 2880N(N-1)(N-2)(N-3).
\end{equation}
\\
Sending $N \rightarrow -N$, yields the $Sp(N)$ result.


\section{Examples of multi-trace operators}

In this appendix, we give two examples of our formula (\ref{eq:26}). Recall from \cite{Kemp:2014apa}, we calculated (with the normalisation of (\ref{eq:2})) 

\begin{eqnarray}
	O_{\ytableausetup{boxsize=0.30em}\ydiagram[]{4,4}\,,(\,\ytableausetup{boxsize=0.30em}\ydiagram[]{2,2}\,,\,\ytableausetup{boxsize=0.30em}\ydiagram[]{2,2}\,)} \!\!\!\! &=& \!\!\!\! \sqrt{30}\Big( 2\mrn{Tr}(ZY)^{2} + 2\mrn{Tr}(ZY\!ZY) + \mrn{Tr}(Z^{2})\mrn{Tr}(Y^{2}) + 4\mrn{Tr}(Z^{2}Y^{2}) \Big)\\
	O_{\ytableausetup{boxsize=0.30em}\ydiagram[]{2,2,2,2}\,,(\,\ytableausetup{boxsize=0.30em}\ydiagram[]{2,2}\,,\,\ytableausetup{boxsize=0.30em}\ydiagram[]{2,2}\,)} \!\!\!\! &=& \!\!\!\! \sqrt{6}\Big( 2\mrn{Tr}(ZY)^{2} + 2\mrn{Tr}(ZY\!ZY) + 3\mrn{Tr}(Z^{2})\mrn{Tr}(Y^{2}) - 12\mrn{Tr}(Z^{2}Y^{2}) \Big)\\
	O_{\ytableausetup{boxsize=0.30em}\ydiagram[]{2,2,2,2}\,,(\,\ytableausetup{boxsize=0.30em}\ydiagram[]{1,1,1,1}\,,\,\ytableausetup{boxsize=0.30em}\ydiagram[]{1,1,1,1}\,)} \!\!\!\! &=& \!\!\!\! 2\sqrt{30}\Big( \mrn{Tr}(ZY)^{2} - 2\mrn{Tr}(ZY\!ZY) \Big)\\
	O_{\ytableausetup{boxsize=0.30em}\ydiagram[]{4,2,2}\,,(\,\ytableausetup{boxsize=0.30em}\ydiagram[]{2,2}\,,\,\ytableausetup{boxsize=0.30em}\ydiagram[]{2,2}\,)} \!\!\!\! &=& \!\!\!\! 2\sqrt{6}\Big( -\mrn{Tr}(ZY)^{2} - \mrn{Tr}(ZY\!ZY) + \mrn{Tr}(Z^{2})\mrn{Tr}(Y^{2}) + \mrn{Tr}(Z^{2}Y^{2}) \Big)
\end{eqnarray}
\\
Let $\sigma=(1,3,5)(4,8,6,7)$. The dual restricted characters evaluated to

\begin{eqnarray}
\label{eq:qq1}
	\chi^{\ytableausetup{boxsize=0.30em}\ydiagram[]{4,4}\,(\,\ytableausetup{boxsize=0.30em}\ydiagram[]{2,2}\,,\,\ytableausetup{boxsize=0.30em}\ydiagram[]{2,2}\,)}_{SO(N)}(\sigma) &=& \frac{1}{6\sqrt{30}}\\
\label{eq:qq2}
	\chi^{\ytableausetup{boxsize=0.30em}\ydiagram[]{2,2,2,2}\,(\,\ytableausetup{boxsize=0.30em}\ydiagram[]{2,2}\,,\,\ytableausetup{boxsize=0.30em}\ydiagram[]{2,2}\,)}_{SO(N)}(\sigma) &=& \frac{1}{30\sqrt{6}}\\
\label{eq:qq3}
	\chi^{\ytableausetup{boxsize=0.30em}\ydiagram[]{2,2,2,2}\,(\,\ytableausetup{boxsize=0.30em}\ydiagram[]{1,1,1,1}\,,\,\ytableausetup{boxsize=0.30em}\ydiagram[]{1,1,1,1}\,)}_{SO(N)}(\sigma) &=& \frac{1}{6\sqrt{30}}\\
\label{eq:qq4}
	\chi^{\ytableausetup{boxsize=0.30em}\ydiagram[]{4,2,2}\,(\,\ytableausetup{boxsize=0.30em}\ydiagram[]{2,2}\,,\,\ytableausetup{boxsize=0.30em}\ydiagram[]{2,2}\,)}_{SO(N)}(\sigma) &=& -\frac{1}{15}\sqrt{\frac{2}{3}}
\end{eqnarray}
\\
Then adding the 4 Schurs with these coefficients, we found

\begin{equation}
\label{eq:qq5}
	\frac{1}{6\sqrt{30}}O_{\ytableausetup{boxsize=0.30em}\ydiagram[]{4,4}\,(\,\ytableausetup{boxsize=0.30em}\ydiagram[]{2,2}\,,\,\ytableausetup{boxsize=0.30em}\ydiagram[]{2,2}\,)}^{SO(N)} +  \frac{1}{30\sqrt{6}}O_{\ytableausetup{boxsize=0.30em}\ydiagram[]{2,2,2,2}\,(\,\ytableausetup{boxsize=0.30em}\ydiagram[]{2,2}\,,\,\ytableausetup{boxsize=0.30em}\ydiagram[]{2,2}\,)}^{SO(N)} + \frac{1}{6\sqrt{30}}O_{\ytableausetup{boxsize=0.30em}\ydiagram[]{2,2,2,2}\,(\,\ytableausetup{boxsize=0.30em}\ydiagram[]{1,1,1,1}\,,\,\ytableausetup{boxsize=0.30em}\ydiagram[]{1,1,1,1}\,)}^{SO(N)} - \frac{1}{15}\sqrt{\frac{2}{3}}O_{\ytableausetup{boxsize=0.30em}\ydiagram[]{4,2,2}\,(\,\ytableausetup{boxsize=0.30em}\ydiagram[]{2,2}\,,\,\ytableausetup{boxsize=0.30em}\ydiagram[]{2,2}\,)}^{SO(N)} = \mrn{Tr}(ZY)^{2}
\end{equation}
\\
We then calculated the right-hand-side of (\ref{eq:26}) and found

\begin{equation}
\label{eq:qq6}
	C^{4\nu}_{I}(\sigma)^{I}_{J}\big( Z^{\otimes 4} \otimes Y^{\otimes 4} \big) =  \mrn{Tr}(ZY)^{2}
\end{equation}
\\
For one more example, Let $\sigma = (3,4,7,8)$. The dual restricted characters evaluated to

\begin{eqnarray}
\label{eq:qq7}
	\chi^{\ytableausetup{boxsize=0.30em}\ydiagram[]{4,4}\,(\,\ytableausetup{boxsize=0.30em}\ydiagram[]{2,2}\,,\,\ytableausetup{boxsize=0.30em}\ydiagram[]{2,2}\,)}_{SO(N)}(\sigma) &=& \frac{1}{12\sqrt{30}}\\
\label{eq:qq8}
	\chi^{\ytableausetup{boxsize=0.30em}\ydiagram[]{2,2,2,2}\,(\,\ytableausetup{boxsize=0.30em}\ydiagram[]{2,2}\,,\,\ytableausetup{boxsize=0.30em}\ydiagram[]{2,2}\,)}_{SO(N)}(\sigma) &=& \frac{-1}{20\sqrt{6}}\\
\label{eq:qq9}
	\chi^{\ytableausetup{boxsize=0.30em}\ydiagram[]{2,2,2,2}\,(\,\ytableausetup{boxsize=0.30em}\ydiagram[]{1,1,1,1}\,,\,\ytableausetup{boxsize=0.30em}\ydiagram[]{1,1,1,1}\,)}_{SO(N)}(\sigma) &=& 0\\
\label{eq:qq10}
	\chi^{\ytableausetup{boxsize=0.30em}\ydiagram[]{4,2,2}\,(\,\ytableausetup{boxsize=0.30em}\ydiagram[]{2,2}\,,\,\ytableausetup{boxsize=0.30em}\ydiagram[]{2,2}\,)}_{SO(N)}(\sigma) &=& \frac{1}{30\sqrt{6}}
\end{eqnarray}
\\
Adding the 4 Schurs with these coefficients, we found

\begin{equation}
\label{eq:qq11}
	 \frac{1}{12\sqrt{30}}O_{\ytableausetup{boxsize=0.30em}\ydiagram[]{4,4}\,(\,\ytableausetup{boxsize=0.30em}\ydiagram[]{2,2}\,,\,\ytableausetup{boxsize=0.30em}\ydiagram[]{2,2}\,)}^{SO(N)} - \frac{1}{20\sqrt{6}}O_{\ytableausetup{boxsize=0.30em}\ydiagram[]{2,2,2,2}\,(\,\ytableausetup{boxsize=0.30em}\ydiagram[]{2,2}\,,\,\ytableausetup{boxsize=0.30em}\ydiagram[]{2,2}\,)}^{SO(N)} + \frac{1}{30\sqrt{6}}O_{\ytableausetup{boxsize=0.30em}\ydiagram[]{4,2,2}\,(\,\ytableausetup{boxsize=0.30em}\ydiagram[]{2,2}\,,\,\ytableausetup{boxsize=0.30em}\ydiagram[]{2,2}\,)}^{SO(N)} = \mrn{Tr}(Z^{2}Y^{2})
\end{equation}
\\
We then calculated the right-hand-side of (\ref{eq:26}) and found

\begin{equation}
\label{eq:qq12}
	C^{4\nu}_{I}(\sigma)^{I}_{J}\big( Z^{\otimes 4} \otimes Y^{\otimes 4} \big) =  \mrn{Tr}(Z^{2}Y^{2})
\end{equation}


\section{Example of the product rule}

In this appendix, we give one simple example of our product rule in (\ref{eq:productrule}). We try evaluate the following product: $O_{\ytableausetup{boxsize=0.30em}\ydiagram[]{2,2},\,(\,\ytableausetup{boxsize=0.30em}\ydiagram[]{1,1}\,,\,\ytableausetup{boxsize=0.30em}\ydiagram[]{1,1}\,)}(Z,Y)O_{\ytableausetup{boxsize=0.30em}\ydiagram[]{2,2},\,(\,\ytableausetup{boxsize=0.30em}\ydiagram[]{1,1}\,,\,\ytableausetup{boxsize=0.30em}\ydiagram[]{1,1}\,)}(Z,Y)$. For the definition in (\ref{eq:Newdefintermsofdelta}),

\begin{equation}
\label{eq:Apppr1}
	O_{\ytableausetup{boxsize=0.30em}\ydiagram[]{2,2},\,(\,\ytableausetup{boxsize=0.30em}\ydiagram[]{1,1}\,,\,\ytableausetup{boxsize=0.30em}\ydiagram[]{1,1}\,)}(Z,Y) = -\sqrt{6}\mrn{Tr}(ZY)
\end{equation}
\\
This means that

\begin{equation}
\label{eq:Apppr2}
	O_{\ytableausetup{boxsize=0.30em}\ydiagram[]{2,2},\,(\,\ytableausetup{boxsize=0.30em}\ydiagram[]{1,1}\,,\,\ytableausetup{boxsize=0.30em}\ydiagram[]{1,1}\,)}(Z,Y)O_{\ytableausetup{boxsize=0.30em}\ydiagram[]{2,2},\,(\,\ytableausetup{boxsize=0.30em}\ydiagram[]{1,1}\,,\,\ytableausetup{boxsize=0.30em}\ydiagram[]{1,1}\,)}(Z,Y) = 6\mrn{Tr}(ZY)^{2}
\end{equation}
\\
According to (\ref{eq:productrule}), this product should also be given as a sum over restricted Schurs, each multiplied by restricted Littlewood-Richardson coefficients, corresponding to the labels in (\ref{eq:26.1}). The operators for these labels evaluated to 

\begin{eqnarray}
\label{eq:Apppr3}
	O_{\ytableausetup{boxsize=0.30em}\ydiagram[]{4,4}\,,(\,\ytableausetup{boxsize=0.30em}\ydiagram[]{2,2}\,,\,\ytableausetup{boxsize=0.30em}\ydiagram[]{2,2}\,)} \!\!\!\! &=& \!\!\!\! \sqrt{30}\Big( 2\mrn{Tr}(ZY)^{2} + 2\mrn{Tr}(ZY\!ZY) + \mrn{Tr}(Z^{2})\mrn{Tr}(Y^{2}) + 4\mrn{Tr}(Z^{2}Y^{2}) \Big)\\
\label{eq:Apppr4}
	O_{\ytableausetup{boxsize=0.30em}\ydiagram[]{2,2,2,2}\,,(\,\ytableausetup{boxsize=0.30em}\ydiagram[]{2,2}\,,\,\ytableausetup{boxsize=0.30em}\ydiagram[]{2,2}\,)} \!\!\!\! &=& \!\!\!\! \sqrt{6}\Big( 2\mrn{Tr}(ZY)^{2} + 2\mrn{Tr}(ZY\!ZY) + 3\mrn{Tr}(Z^{2})\mrn{Tr}(Y^{2}) - 12\mrn{Tr}(Z^{2}Y^{2}) \Big)\\
\label{eq:Apppr5}
	O_{\ytableausetup{boxsize=0.30em}\ydiagram[]{2,2,2,2}\,,(\,\ytableausetup{boxsize=0.30em}\ydiagram[]{1,1,1,1}\,,\,\ytableausetup{boxsize=0.30em}\ydiagram[]{1,1,1,1}\,)} \!\!\!\! &=& \!\!\!\! 2\sqrt{30}\Big( \mrn{Tr}(ZY)^{2} - 2\mrn{Tr}(ZY\!ZY) \Big)\\
\label{eq:Apppr6}
	O_{\ytableausetup{boxsize=0.30em}\ydiagram[]{4,2,2}\,,(\,\ytableausetup{boxsize=0.30em}\ydiagram[]{2,2}\,,\,\ytableausetup{boxsize=0.30em}\ydiagram[]{2,2}\,)} \!\!\!\! &=& \!\!\!\! -2\sqrt{6}\Big( -\mrn{Tr}(ZY)^{2} - \mrn{Tr}(ZY\!ZY) +\mrn{Tr}(Z^{2})\mrn{Tr}(Y^{2}) + \mrn{Tr}(Z^{2}Y^{2}) \Big)
\end{eqnarray}
\\
Only the last operator differs from the operators defined in (\ref{eq:2}) and the difference is a minus sign. Here, $n_{1} = m_{1} = n_{2} = m_{2} = 1$, and $q_{1}=q_{2}=2$. The restricted Littlewood-Richardson coefficients evaluated to 

\begin{eqnarray}
\label{eq:Apppr7}
	f^{\ytableausetup{boxsize=0.30em}\ydiagram[]{4,4}\,,(\,\ytableausetup{boxsize=0.30em}\ydiagram[]{2,2}\,,\,\ytableausetup{boxsize=0.30em}\ydiagram[]{2,2}\,)}_{\ytableausetup{boxsize=0.30em}\ydiagram[]{2,2}\,,(\,\ytableausetup{boxsize=0.30em}\ydiagram[]{1,1}\,,\,\ytableausetup{boxsize=0.30em}\ydiagram[]{1,1}\,)\; ;\;\ytableausetup{boxsize=0.30em}\ydiagram[]{2,2}\,,(\,\ytableausetup{boxsize=0.30em}\ydiagram[]{1,1}\,,\,\ytableausetup{boxsize=0.30em}\ydiagram[]{1,1}\,)} &=& \frac{1}{\sqrt{30}}\\
\label{eq:Apppr8}
	f^{\ytableausetup{boxsize=0.30em}\ydiagram[]{2,2,2,2}\,,(\,\ytableausetup{boxsize=0.30em}\ydiagram[]{2,2}\,,\,\ytableausetup{boxsize=0.30em}\ydiagram[]{2,2}\,)}_{\ytableausetup{boxsize=0.30em}\ydiagram[]{2,2}\,,(\,\ytableausetup{boxsize=0.30em}\ydiagram[]{1,1}\,,\,\ytableausetup{boxsize=0.30em}\ydiagram[]{1,1}\,)\; ;\;\ytableausetup{boxsize=0.30em}\ydiagram[]{2,2}\,,(\,\ytableausetup{boxsize=0.30em}\ydiagram[]{1,1}\,,\,\ytableausetup{boxsize=0.30em}\ydiagram[]{1,1}\,)} &=& \frac{1}{5\sqrt{6}}\\
\label{eq:Apppr9}
	f^{\ytableausetup{boxsize=0.30em}\ydiagram[]{2,2,2,2}\,,(\,\ytableausetup{boxsize=0.30em}\ydiagram[]{1,1,1,1}\,,\,\ytableausetup{boxsize=0.30em}\ydiagram[]{1,1,1,1}\,)}_{\ytableausetup{boxsize=0.30em}\ydiagram[]{2,2}\,,(\,\ytableausetup{boxsize=0.30em}\ydiagram[]{1,1}\,,\,\ytableausetup{boxsize=0.30em}\ydiagram[]{1,1}\,)\; ;\;\ytableausetup{boxsize=0.30em}\ydiagram[]{2,2}\,,(\,\ytableausetup{boxsize=0.30em}\ydiagram[]{1,1}\,,\,\ytableausetup{boxsize=0.30em}\ydiagram[]{1,1}\,)} &=& \frac{1}{\sqrt{30}}\\
\label{eq:Apppr10}
	f^{\ytableausetup{boxsize=0.30em}\ydiagram[]{4,2,2}\,,(\,\ytableausetup{boxsize=0.30em}\ydiagram[]{2,2}\,,\,\ytableausetup{boxsize=0.30em}\ydiagram[]{2,2}\,)}_{\ytableausetup{boxsize=0.30em}\ydiagram[]{2,2}\,,(\,\ytableausetup{boxsize=0.30em}\ydiagram[]{1,1}\,,\,\ytableausetup{boxsize=0.30em}\ydiagram[]{1,1}\,)\; ;\;\ytableausetup{boxsize=0.30em}\ydiagram[]{2,2}\,,(\,\ytableausetup{boxsize=0.30em}\ydiagram[]{1,1}\,,\,\ytableausetup{boxsize=0.30em}\ydiagram[]{1,1}\,)} &=& \frac{2}{5}\sqrt{\frac{2}{3}}
\end{eqnarray}
\\
In the $f$'s above, there were two sums over the permutation group $S_{4}$. The first sum was over permutations that permuted $1,2,5,6$ amongst themselves, and the second sum was over permutations that permuted $3,4,7,8$ amongst themselves. For the coefficients in (\ref{eq:Apppr7}) to (\ref{eq:Apppr10}) and operators (\ref{eq:Apppr3}) to (\ref{eq:Apppr6}), we found

\begin{eqnarray}
\label{eq:Apppr11}
	&&f^{\ytableausetup{boxsize=0.30em}\ydiagram[]{4,4}\,,(\,\ytableausetup{boxsize=0.30em}\ydiagram[]{2,2}\,,\,\ytableausetup{boxsize=0.30em}\ydiagram[]{2,2}\,)}_{\ytableausetup{boxsize=0.30em}\ydiagram[]{2,2}\,,(\,\ytableausetup{boxsize=0.30em}\ydiagram[]{1,1}\,,\,\ytableausetup{boxsize=0.30em}\ydiagram[]{1,1}\,)\; ;\;\ytableausetup{boxsize=0.30em}\ydiagram[]{2,2}\,,(\,\ytableausetup{boxsize=0.30em}\ydiagram[]{1,1}\,,\,\ytableausetup{boxsize=0.30em}\ydiagram[]{1,1}\,)}O_{\ytableausetup{boxsize=0.30em}\ydiagram[]{4,4}\,,(\,\ytableausetup{boxsize=0.30em}\ydiagram[]{2,2}\,,\,\ytableausetup{boxsize=0.30em}\ydiagram[]{2,2}\,)}(Y,Z) + f^{\ytableausetup{boxsize=0.30em}\ydiagram[]{2,2,2,2}\,,(\,\ytableausetup{boxsize=0.30em}\ydiagram[]{2,2}\,,\,\ytableausetup{boxsize=0.30em}\ydiagram[]{2,2}\,)}_{\ytableausetup{boxsize=0.30em}\ydiagram[]{2,2}\,,(\,\ytableausetup{boxsize=0.30em}\ydiagram[]{1,1}\,,\,\ytableausetup{boxsize=0.30em}\ydiagram[]{1,1}\,)\; ;\;\ytableausetup{boxsize=0.30em}\ydiagram[]{2,2}\,,(\,\ytableausetup{boxsize=0.30em}\ydiagram[]{1,1}\,,\,\ytableausetup{boxsize=0.30em}\ydiagram[]{1,1}\,)}O_{\ytableausetup{boxsize=0.30em}\ydiagram[]{2,2,2,2}\,,(\,\ytableausetup{boxsize=0.30em}\ydiagram[]{2,2}\,,\,\ytableausetup{boxsize=0.30em}\ydiagram[]{2,2}\,)} + f^{\ytableausetup{boxsize=0.30em}\ydiagram[]{2,2,2,2}\,,(\,\ytableausetup{boxsize=0.30em}\ydiagram[]{1,1,1,1}\,,\,\ytableausetup{boxsize=0.30em}\ydiagram[]{1,1,1,1}\,)}_{\ytableausetup{boxsize=0.30em}\ydiagram[]{2,2}\,,(\,\ytableausetup{boxsize=0.30em}\ydiagram[]{1,1}\,,\,\ytableausetup{boxsize=0.30em}\ydiagram[]{1,1}\,)\; ;\;\ytableausetup{boxsize=0.30em}\ydiagram[]{2,2}\,,(\,\ytableausetup{boxsize=0.30em}\ydiagram[]{1,1}\,,\,\ytableausetup{boxsize=0.30em}\ydiagram[]{1,1}\,)}O_{\ytableausetup{boxsize=0.30em}\ydiagram[]{2,2,2,2}\,,(\,\ytableausetup{boxsize=0.30em}\ydiagram[]{1,1,1,1}\,,\,\ytableausetup{boxsize=0.30em}\ydiagram[]{1,1,1,1}\,)} \nonumber \\
	&& +\;f^{\ytableausetup{boxsize=0.30em}\ydiagram[]{4,2,2}\,,(\,\ytableausetup{boxsize=0.30em}\ydiagram[]{2,2}\,,\,\ytableausetup{boxsize=0.30em}\ydiagram[]{2,2}\,)}_{\ytableausetup{boxsize=0.30em}\ydiagram[]{2,2}\,,(\,\ytableausetup{boxsize=0.30em}\ydiagram[]{1,1}\,,\,\ytableausetup{boxsize=0.30em}\ydiagram[]{1,1}\,)\; ;\;\ytableausetup{boxsize=0.30em}\ydiagram[]{2,2}\,,(\,\ytableausetup{boxsize=0.30em}\ydiagram[]{1,1}\,,\,\ytableausetup{boxsize=0.30em}\ydiagram[]{1,1}\,)}  O_{\ytableausetup{boxsize=0.30em}\ydiagram[]{4,2,2}\,,(\,\ytableausetup{boxsize=0.30em}\ydiagram[]{2,2}\,,\,\ytableausetup{boxsize=0.30em}\ydiagram[]{2,2}\,)} = 6\mrn{Tr}(ZY)^{2}
\end{eqnarray}
\\
precisely as expected.
\bibliographystyle{unsrt}

\bibliography{Refs}

\begin{thebibliography}{10}

\bibitem{Kemp:2014apa}
Garreth Kemp.
\newblock {SO(N) restricted Schur polynomials}.
\newblock arXiv 1405.7017, 2014.

\bibitem{GGWSA1}
Robert de~Mello~Koch, Jelena Smolic, and Milena Smolic.
\newblock {Giant Gravitons - with Strings Attached (I)}.
\newblock {\em JHEP}, 0706:074, 2007.

\bibitem{GGWSA2}
Robert de~Mello~Koch, Jelena Smolic, and Milena Smolic.
\newblock {Giant Gravitons - with Strings Attached (II)}.
\newblock {\em JHEP}, 0709:049, 2007.

\bibitem{GGWSA3}
David Bekker, Robert de~Mello~Koch, and Michael Stephanou.
\newblock {Giant Gravitons - with Strings Attached. III.}
\newblock {\em JHEP}, 0802:029, 2008.

\bibitem{Jevicki}
Steve Corley, Antal Jevicki, and Sanjaye Ramgoolam.
\newblock {Exact correlators of giant gravitons from dual N=4 SYM theory}.
\newblock {\em Adv.Theor.Math.Phys.}, 5:809--839, 2002.

\bibitem{Invasion}
John McGreevy, Leonard Susskind, and Nicolaos Toumbas.
\newblock {Invasion of the giant gravitons from Anti-de Sitter space}.
\newblock {\em JHEP}, 0006:008, 2000.

\bibitem{Hashimoto:2000zp}
Akikazu Hashimoto, Shinji Hirano, and N.~Itzhaki.
\newblock {Large branes in AdS and their field theory dual}.
\newblock {\em JHEP}, 0008:051, 2000.

\bibitem{Meyers}
Marcus~T. Grisaru, Robert~C. Myers, and Oyvind Tafjord.
\newblock {SUSY and goliath}.
\newblock {\em JHEP}, 0008:040, 2000.

\bibitem{Emergent}
Robert de~Mello Koch and Jeff Murugan.
\newblock {Emergent Spacetime}.
\newblock pages 164--184, 2009.

\bibitem{XYZmatrices}
Robert de~Mello Koch, Badr Awad~Elseid Mohammed, and Stephanie Smith.
\newblock {Nonplanar Integrability: Beyond the SU(2) Sector}.
\newblock {\em Int.J.Mod.Phys.}, A26:4553--4583, 2011.

\bibitem{Fermions}
Robert de~Mello~Koch, Pablo Diaz, and Nkululeko Nokwara.
\newblock {Restricted Schur Polynomials for Fermions and integrability in the
  su(2|3) sector}.
\newblock {\em JHEP}, 1303:173, 2013.

\bibitem{deMelloKoch:2011vn}
Robert de~Mello~Koch, Pablo Diaz, and Hesam Soltanpanahi.
\newblock {Non-planar Anomalous Dimensions in the sl(2) Sector}.
\newblock {\em Phys.Lett.}, B713:509--513, 2012.

\bibitem{Collins}
Storm Collins.
\newblock {Restricted Schur Polynomials and Finite N Counting}.
\newblock {\em Phys.Rev.}, D79:026002, 2009.

\bibitem{EMC}
Rajsekhar Bhattacharyya, Storm Collins, and Robert de~Mello Koch.
\newblock {Exact Multi-Matrix Correlators}.
\newblock {\em JHEP}, 0803:044, 2008.

\bibitem{Bhattacharyya:2008xy}
Rajsekhar Bhattacharyya, Robert de~Mello~Koch, and Michael Stephanou.
\newblock {Exact Multi-Restricted Schur Polynomial Correlators}.
\newblock {\em JHEP}, 0806:101, 2008.

\bibitem{Kimura:2007wy}
Yusuke Kimura and Sanjaye Ramgoolam.
\newblock {Branes, anti-branes and brauer algebras in gauge-gravity duality}.
\newblock {\em JHEP}, 0711:078, 2007.

\bibitem{Brown:2007xh}
Thomas~William Brown, P.J. Heslop, and S.~Ramgoolam.
\newblock {Diagonal multi-matrix correlators and BPS operators in N=4 SYM}.
\newblock {\em JHEP}, 0802:030, 2008.

\bibitem{Weakcoup}
Jurgis Pasukonis and Sanjaye Ramgoolam.
\newblock {From counting to construction of BPS states in N=4 SYM}.
\newblock {\em JHEP}, 1102:078, 2011.

\bibitem{Koch:2010gp}
Robert de~Mello Koch, Grant Mashile, and Nicholas Park.
\newblock {Emergent Threebrane Lattices}.
\newblock {\em Phys.Rev.}, D81:106009, 2010.

\bibitem{DeComarmond:2010ie}
Vincent De~Comarmond, Robert de~Mello~Koch, and Katherine Jefferies.
\newblock {Surprisingly Simple Spectra}.
\newblock {\em JHEP}, 1102:006, 2011.

\bibitem{Carlson:2011hy}
Warren Carlson, Robert de~Mello Koch, and Hai Lin.
\newblock {Nonplanar Integrability}.
\newblock {\em JHEP}, 1103:105, 2011.

\bibitem{GGO}
Robert de~Mello Koch, Matthias Dessein, Dimitrios Giataganas, and Christopher
  Mathwin.
\newblock {Giant Graviton Oscillators}.
\newblock {\em JHEP}, 1110:009, 2011.

\bibitem{Spring}
Robert de~Mello~Koch, Garreth Kemp, and Stephanie Smith.
\newblock {From Large N Nonplanar Anomalous Dimensions to Open Spring Theory}.
\newblock {\em Phys.Lett.}, B711:398--403, 2012.

\bibitem{deMelloKoch:2012ck}
Robert de~Mello~Koch and Sanjaye Ramgoolam.
\newblock {A double coset ansatz for integrability in AdS/CFT}.
\newblock {\em JHEP}, 1206:083, 2012.

\bibitem{twoloop}
Robert de~Mello~Koch, Garreth Kemp, Badr Awad~Elseid Mohammed, and Stephanie
  Smith.
\newblock {Nonplanar integrability at two loops}.
\newblock {\em JHEP}, 1210:144, 2012.

\bibitem{SON1}
Pawel Caputa, Robert de~Mello~Koch, and Pablo Diaz.
\newblock {A basis for large operators in N=4 SYM with orthogonal gauge group}.
\newblock {\em JHEP}, 1303:041, 2013.

\bibitem{SON2}
Pawel Caputa, Robert de~Mello Koch, and Pablo Diaz.
\newblock {Operators, Correlators and Free Fermions for SO(N) and Sp(N)}.
\newblock {\em JHEP}, 1306:018, 2013.

\bibitem{Diaz}
Pablo Diaz.
\newblock {Orthogonal Schurs for Classical Gauge Groups}.
\newblock {\em JHEP}, 1310:228, 2013.

\bibitem{SpecProbSON}
Pawel Caputa, Charlotte Kristjansen, and Konstantinos Zoubos.
\newblock {On the spectral problem of N=4 SYM with orthogonal or symplectic
  gauge group}.
\newblock {\em JHEP}, 1010:082, 2010.

\bibitem{Cicuta:1982fu}
G.M. Cicuta.
\newblock {Topological Expansion for SO($N$) and Sp(2n) Gauge Theories}.
\newblock {\em Lett.Nuovo Cim.}, 35:87, 1982.

\bibitem{Dolan2}
F.A. Dolan.
\newblock {Counting BPS operators in N=4 SYM}.
\newblock {\em Nucl.Phys.}, B790:432--464, 2008.

\bibitem{Aharony:2003sx}
Ofer Aharony, Joseph Marsano, Shiraz Minwalla, Kyriakos Papadodimas, and Mark
  Van~Raamsdonk.
\newblock {The Hagedorn - deconfinement phase transition in weakly coupled
  large N gauge theories}.
\newblock {\em Adv.Theor.Math.Phys.}, 8:603--696, 2004.

\bibitem{Dolan1}
F.A. Dolan and H.~Osborn.
\newblock {Applications of the Superconformal Index for Protected Operators and
  q-Hypergeometric Identities to N=1 Dual Theories}.
\newblock {\em Nucl.Phys.}, B818:137--178, 2009.

\bibitem{SymmFunc}
I.G. MacDonald.
\newblock {\em {Symmetric Functions and Hall Polynomials}}.
\newblock Oxford Mathematical Monographs, 1995.

\bibitem{2010arXiv1011.4734V}
V.~{Venkateswaran}.
\newblock {Vanishing integrals for Hall-Littlewood polynomials}.
\newblock {\em ArXiv e-prints}, November 2010.

\bibitem{Vanishing}
E.M. Rains and Monica Vazirani.
\newblock {Vanishing integrals of MacDonald Koornwinder polynomials}.
\newblock {\em Transformation Groups}, Volume 12:725--759, 2007.

\bibitem{Ivanov}
V.N. Ivanov.
\newblock {bispherical functions on the symmetric group associated with the
  hyperoctahedral group}.
\newblock {\em Journal of Mathematical Sciences}, Vol. 96, No. 5:3505--3516,
  1999.

\bibitem{Quiver}
Jurgis Pasukonis and Sanjaye Ramgoolam.
\newblock {Quivers as Calculators: Counting, Correlators and Riemann Surfaces}.
\newblock {\em JHEP}, 1304:094, 2013.

\end{thebibliography}

\end{document}